\documentclass[12pt,a4paper]{JHEP3}
\preprint{Cavendish-HEP-12/15 \\
DESY 12-163\\
IPMU12-0185 \\
KEK-TH 1580\\
}
\usepackage{cite}
\usepackage{amssymb}
\usepackage{amsmath}
\usepackage{enumerate}
\usepackage{slashed}
\usepackage{graphicx}
\newcommand{\beq}{\begin{equation}}
\newcommand{\eeq}{\end{equation}}
\newcommand{\beqn}{\begin{eqnarray}}
\newcommand{\eeqn}{\end{eqnarray}}
\newcommand{\bmat}{\begin{pmatrix}}
\newcommand{\emat}{\end{pmatrix}}

\def\vp{\mathbf{p}}
\def\vq{\mathbf{q}}
\def\vr{\mathbf{r}}
\def\vpm{\slashed{\vp}}

\title{Reconstruction of Higgs bosons in the di-tau channel via 3-prong decay}

\author{Ben Gripaios$^1$, Keiko Nagao$^2$, Mihoko Nojiri$^{2,3,4}$, Kazuki Sakurai$^5$, Bryan Webber$^1$\\
$^1$Cavendish Laboratory, University of Cambridge,  J.J.~Thomson Avenue,  Cambridge, UK\\
$^2$KEK Theory Center, Tsukuba, Ibaraki 305-0801, Japan\\ 
$^3$The Graduate University for Advanced Studies (Sokendai) 
IPNS, KEK,  Tsukuba, Ibaraki, 305-0801, Japan \\
$^4$Kavli Institute of the Physics and Mathematics of the Universe (Kavli IPMU)\\
$^5$DESY, Notkestrasse 85, D-22607 Hamburg, Germany\\
Email: \email{gripaios@hep.phy.cam.ac.uk, knagao@post.kek.jp, nojiri@post.kek.jp,
  kazuki.sakurai@desy.de, webber@hep.phy.cam.ac.uk}}
  
\abstract{
We propose a method for reconstructing the mass of a particle, such as
the Higgs boson, decaying into a pair of $\tau$ leptons, of which one
subsequently undergoes a 3-prong decay. The kinematics is solved using
information
from the visible decay products, the missing transverse momentum, and
the 3-prong $\tau$ decay vertex, with the detector resolution taken
into account using a likelihood method. 
The method is shown to give  good discrimination between a 125 GeV Higgs boson
signal and the dominant backgrounds, such as $Z^0$ decays to $\tau
\tau$  and $W^\pm$
plus jets production. As a result, we find an improvement, compared to
existing methods for this channel, in the discovery potential, as well
as in measurements of the Higgs boson mass and production cross section times
branching ratio.
}   


\begin{document}
\maketitle
\flushbottom


\section{Introduction}
The Standard Model Higgs boson -- or something like it -- has been
found \cite{:2012gk,:2012gu} and the race is now on to determine its
detailed properties. 
Of particular interest is the $\tau \tau$ decay
channel, where, despite sensitivity to a signal comparable to that predicted in
the Standard Model with $m_h \simeq 125$ GeV, 
only weak evidence somewhat below the Standard Model expectation has been seen~\cite{Einsweiler:HCP, Paus:HCP}.

This may be evidence for physics beyond the Standard Model or it may be
a statistical fluctuation. 
We should dearly like to get to the bottom of the mystery, but progress is slowed by a number of
complications in the $h \rightarrow
\tau \tau$ search channel.

One complication is
the problem of triggering and identifying $\tau$ leptons, which decay
in a variety of ways, some of which resemble
common-or-garden QCD jets. A second is that, since a $\tau$ decay always
involves one or more invisible neutrinos, 
it is only possible to reconstruct the resonance in an approximate way.
A third is the problem of
distinguishing $h \rightarrow
\tau \tau$ signal events from the dominant backgrounds, namely 
$Z \rightarrow
\tau \tau$ decays (which, moreover, lie nearby in $\tau \tau$ invariant mass)
and production of QCD jets, with or without a $W$ boson, in which
other leptons or jets are mistakenly identified as $\tau$ leptons.

The third issue, of distinguishing the signal from backgrounds, is 
compounded by the second issue, that we are {\em a priori} unable to
reconstruct the  invariant mass of the $\tau \tau$ pair. 
As a result, and as is readily apparent from the data shown in \cite{CMS11}, the signal and backgrounds lie close to each other in distributions of the observables that are currently available to us.
Not only does this
increase the integrated luminosity required to be sure that an
apparent signal is not a mere background
fluctuation, but it also increases our vulnerability to systematic
errors.

To be explicit, imagine a hypothetical limit in which we have
a signal (the Higgs) that consists of a narrow peak in
invariant mass, a background (the $Z$) that is also a narrow
peak, but centred elsewhere, and additional backgrounds (like $W$ plus
jets and QCD) that are approximately flat in invariant mass. The signals and backgrounds are otherwise roughly
indistinguishable in their dynamics. 
Now, if the observables that we have available are uncorrelated with the invariant
mass, then we are essentially reduced to counting events in order to
try to discover the signal and our ability to do so is greatly limited
by the total statistics available. We are, moreover, completely at the
mercy of systematic uncertainties in the overall background
normalization, which we have no way to measure in data.
Even if we were able to make a discovery in this way, we could at most
make one measurement of the signal properties (its overall size) and
here too we would be exposed to the systematic uncertainty in the
background normalization.

Conversely, if we find a way to reconstruct, more or less, the invariant
mass, the first benefit is that we are no longer limited by the
overall statistics, but rather by the number of signal and background
events in a region of invariant mass of our choosing (near 125 GeV
being the obvious choice for $h \rightarrow \tau \tau$).
Moreover, we
now have a clean separation between the signal and background and
indeed between the different backgrounds themselves. 
This opens up the possibility of using the extra information to constrain 
the uncertainties on the background yields and shapes via data-driven techniques.
Indeed, a simple sideband analysis would
suffice, in which the $Z$ background is measured in a `control' region near 90
GeV and the other backgrounds are measured in a control region away
from the peaks
near 90 GeV and 125 GeV. Finally, independent measurements of the signal mass and
cross-section times branching ratio become possible.

Needless to say, the real situation is rather more complicated for $h\rightarrow \tau \tau$ at the LHC, with the current performance falling somewhere between the two extremes of perfect and imperfect mass resolution. Nevertheless, the basic principle remains the same:
the more we are
able to separate the signal and the different background components
from each other, the less we shall find ourselves at the mercy of
statistical and systematic uncertainties.

So, how could we reconstruct something like the $\tau \tau$ invariant
mass? Several approximate methods or observables have previously been suggested in the literature
(see, for example, \cite{Ellis:1987xu,CMS2,Elagin:2010aw,Gripaios:2011jm,Barr:2011he,Barr:2011si}).
Some of these suffer from being rather poorly correlated with the invariant mass (some provide, for example, only an upper or lower bound on it), while others suffer from the fact that they turn out to be ill-defined for a significant fraction of events, with a consequent loss of statistics. As examples, the collinear approximation used in \cite{Barr:2011he} fails for one in three events, whereas the observable used in \cite{Ellis:1987xu, Barr:2011si} does not exist for a similar fraction of events.

Here we wish to propose yet another method, which differs
significantly in that  we focus on the subset of events in which a $\tau$ lepton undergoes a 3-prong
decay. This implies an immediate disadvantage in the form
of a reduced number of signal events overall for a given integrated luminosity: a
$\tau$ lepton has a branching ratio of 15 \% for a
3-prong decay (of which 9.3 \% are to $\pi^-\pi^+\pi^-\nu_\tau$
and 4.6 \% are to $\pi^-\pi^+\pi^- \pi^0 \nu_\tau$) \cite{Amsler:2008zzb}, meaning that only 28 \% of di-$\tau$ events feature at least one
3-prong decay.
However, the hope is that this disadvantage is more than compensated by the advantages.

These advantages all stem from the fact that the presence of a 3-prong
decay allows us to
reconstruct the $\tau \tau$ invariant mass, if the invariant mass of
the neutrino or neutrinos from the other $\tau$ decay is known. 
Thus, for hadronic decays of the other $\tau$, we can fully reconstruct events (up to a
discrete ambiguity and in the absence
of detector mismeasurements, both of which we shall deal with below);
for leptonic decays of the other $\tau$, we are able to partially
reconstruct events. As a result we hope to benefit from a reduced
exposure to statistical and systematic uncertainties as argued above.

The extra kinematic information needed to reconstruct comes from the
location of the secondary (3-prong $\tau$-decay) vertex: if one can
measure with reasonable accuracy the impact parameter of each of the
three charged tracks, defined as the shortest distance between the
track and the primary vertex, then the intersection of these impact parameters
gives the location of the secondary vertex.\footnote{The fact that
  there are three intersections means that we have an indication of
  the quality of the vertex reconstruction in an individual event. This
  information could, in principle, be fed into the likelihood function
that we shall use to account for detector response, but we do not do
so here.} 

Now, the location of the secondary vertex tells us the direction of
the $\tau$ momentum; the mass-shell constraint for the
$\tau$ then allows to reconstruct the magnitude of the $\tau$
momentum, up to a possible two-fold ambiguity. Given the measured
missing transverse momentum, we are then able to reconstruct the
momentum of the other $\tau$ (up to a further possible two-fold ambiguity), provided we know the invariant mass of
the neutrino(s) produced in the other $\tau$ decay.

The reconstruction process just described can only be expected to work
if things are well measured. For example, if they are not, we may end
up with no real solutions to the kinematic constraints.
We account for this by
defining an {\em ad hoc} likelihood function in which we convolute
the observed quantities with a function parameterizing the detector
response. The maximum of this likelihood function is an event
observable (albeit one with an obscure definition) and it is this
observable that we propose to use for signal discrimination. 
The fact that we invoke a likelihood function also allows to deal with
the unknown invariant mass of the two neutrinos produced in a leptonic
$\tau$ decay: we marginalize
with respect to the unknown invariant mass,
including the matrix element for the $\tau$ decay.

Yet another advantage of focussing on 3-prong decays is that the fake backgrounds (coming from, e.g. W+ jets and QCD) will be reduced, as jets and other leptons are presumably less likely to fake a 3-prong decay (with a reconstructed secondary vertex) than they are to fake a generic hadronic tau or leptonic tau decay.\footnote{
Unfortunately, it is not possible for us to reliably
estimate the size of this effect, since neither the full details of the
experimental $\tau$ reconstruction algorithm nor the resulting
efficiencies or fake rates for 3-prong decays are public.}


The outline of the paper is as follows. In the next Section, we describe the algebraic details of the
reconstruction procedure. In Section \ref{sec:results}, we present the
results of our numerical simulations and in Section
\ref{sec:conclusions}, we draw our conclusions.
\section{The method \label{sec:method}}
As described in the introduction, events are reconstructed using the decay
vertex information.  If one of the $\tau$ leptons decays to a 3-prong
hadronic system and a $\tau$ neutrino, then the displacement $\vr$ from the primary interaction point to the $\tau$ decay vertex
should be measurable with useful precision.\footnote{The ATLAS
CSC~\cite{Aad:2009wy} estimates a resolution of $0.6$ and $0.01$ mm in the
directions parallel and perpendicular to the displacement,
respectively.}

Let us begin by considering the limit of perfect detector resolution. Denoting the energy, momentum and mass of the 3-prong decay
products by $E_j$, $\vp_j$ and $m_j$, respectively, and the angle
between $\vr$ and $\vp_j$ by $\theta$,  the momentum of that
$\tau$ lepton can be reconstructed, with a twofold ambiguity,
as $\vp_\tau = p_\tau\vr/|\vr|$, where (neglecting the neutrino mass)
\beq \label{eq:reconptau}
p_\tau=\frac{(m_\tau^2+m_j^2)p_j\cos\theta\pm E_j
\sqrt{(m_\tau^2-m_j^2)^2-4m_\tau^2p_j^2\sin^2\theta}}
{2(m_j^2+p_j^2\sin^2\theta)}\,.
\eeq

The other $\tau$ lepton may decay either hadronically or leptonically,
into a visible system $j'$ (a hadronic jet or a charged lepton) and an
invisible system $i'$ (a $\tau$ neutrino or a pair of neutrinos).  The
transverse momentum of the invisible system is found from the missing
transverse momentum, $\vpm_T$, and the reconstructed momentum
of the first $\tau$ via
\beq
\vp_{Ti'} = \vpm_T+\vp_{Tj}-\vp_{T\tau}\,.
\label{eq:pmiss}
\eeq
Given the invariant mass of the invisible system, $m_{i'}=m_\nu=0$ for
a hadronic and $m_{i'}=m_{\nu\nu}\geq 0$ for a leptonic decay, one can then solve for
the invisible longitudinal momentum, again with a twofold ambiguity:
\beq \label{eq:reconpl}
p_{L {i'}} = \frac 1{\mu_{j'}}\left(\alpha\,p_{Lj'}\pm E_{j'}
\sqrt{\alpha^2-\mu_{i'} \mu_{j'}}\right)\,,
\eeq
where
\beq
\mu_{i'} = p_{T {i'}}^2+m_{i'}^2\,,\;\;
\mu_{j'} = p_{Tj'}^2+m_{j'}^2\,,\;\;
\alpha=\frac 12 (m_\tau^2-m_{j'}^2-m_{i'}^2)  +  \vp_{T {i'}}\cdot\vp_{Tj'}\,.
\eeq
The momentum of the second $\tau$ can now be reconstructed as
$\vp_{\tau'} = \vp_{i'}+\vp_{j'}$, and hence the invariant mass of the
$\tau\tau$ system, $m_{\tau \tau}$ is determined, up to a fourfold ambiguity.

Now consider a real detector and let 
$\vq =(\vr,E_j,\vp_j,\vp_j',\vpm_T)$ correspond to the measured
quantities. These do not coincide with
their true values in an event, which we now denote by $\tilde\vq$, but rather are shifted by amounts depending on the
detector resolution, which 
we describe by a response function,
$f(\vq,\tilde\vq)$.  Then the
likelihood, as a function of the true invariant mass
$\tilde{m}_{\tau \tau}$, for an event with measured
quantities $\vq$, may be written as
\beq\label{eq:like}
{\cal L}(\tilde{m}_{\tau \tau}|\vq) = \int
d\tilde\vq\,f(\vq,\tilde\vq)\,{\cal
  M}(\tilde\vq)\,\delta[\tilde{m}_{\tau \tau} - m_{\tau \tau} (\tilde\vq)]
\eeq
where ${\cal M}(\tilde\vq)$ is the matrix-element squared for the decay and
$m_{\tau \tau} (\tilde\vq)$ is the invariant $\tau \tau$ mass
reconstructed from the true quantities $\tilde\vq$ according to the
recipe described above.
Here ${\cal M}(\tilde\vq)$ should also include the jacobian factor relating the
final-state phase space to the quantities $\tilde\vq$. We find, in
most cases, that including these effects gives, at best, a marginal
improvement in the mass resolution. Indeed, some effects 
(such as
the exponential distribution of the $\tau$-decay lifetimes), 
lead to large fluctuations in the likelihood integrand and hence to
large errors in the numerical integration, worsening the mass
resolution.
Thus we do not include these effects, in general. 

There is, however, one such effect which we do include. In the case of
leptonic decay of the second $\tau$,  the matrix elements also depend
on the momenta of the two invisible neutrinos, and the right-hand side
of eq.~(\ref{eq:like}) should include an integration over their phase
space, weighted by the expected distribution of the  $\nu\nu$
invariant mass. This is conveniently expressed as
 $P(m_{\nu\nu}^2)\,d\Phi_{\nu\nu}$, with
\beq
d\Phi_{\nu\nu} = \frac{d^3\vp_{\nu\nu}}{2(2\pi)^3E_{\nu\nu}}
\frac{d\Omega^*}{(4\pi)^3}dm_{\nu\nu}^2\,,
\eeq
where $d\Omega^*$ is the element of solid angle in the $\nu\nu$
centre-of-mass frame and
\beq
P(m_{\nu\nu}^2) = \frac{2}{m_\tau^2}
\left(1-\frac{m_{\nu\nu}^2}{m_\tau^2}\right)^2
\left(1+2\frac{m_{\nu\nu}^2}{m_\tau^2}\right)\,.
\eeq
At each phase-space point, 
the value of $m_{\nu\nu}$ is then used, with this weight, for the reconstruction of the decay. 

In the integral over the delta function in eq.~(\ref{eq:like}),
we include all real
solutions to the equation $\tilde{m}_{\tau \tau} = m_{\tau \tau} (\tilde\vq)$. In contrast to
\cite{Gripaios:2011jm}, complex solutions should be discarded here,
since they cannot correspond to true values for genuine $\tau \tau$
resonance events.
Note, however, that measured values $\vq$ that would correspond to
complex values of $m_{\tau \tau}$ lying close to the real axis if
reconstructed directly, which correspond to real solutions shifted
slightly by detector resolution, will be included in the integration
at neighbouring values of $\tilde\vq$. For fake backgrounds, we often
find that no nearby values of $\tilde\vq$ lead to real solutions,
allowing the event to be rejected.

We perform the integrations in eq.~(\ref{eq:like}) by a Monte Carlo
method similar to that adopted in \cite{Kawagoe:2004rz}, generating
a large number of points $\tilde\vq$ distributed around each measured
point $\vq$ according to a smearing function $f(\vq,\tilde\vq)$ deduced
from detector simulations. 
The jet masses are generated according to certain probabilities that
we describe in the Appendix.
Each real solution $\tilde{m}_{\tau \tau} = m_{\tau \tau} (\tilde\vq)$ is
entered into a histogram with the corresponding weight.   
Because the Monte Carlo method generates
only a finite number of points, all histogram bins are given a small
positive offset, to avoid multiplications by zero.

Since our likelihood function does not encode the matrix element in
its entirety, we cannot expect to be able to make statistical
inferences directly from it in the usual way. Doing so might lead, for
example, to us wrongly rejecting the Standard Model Higgs boson
hypothesis, or obtaining a biased measurement of its mass. 
Instead, we use our {\em ad hoc} likelihood function to define an
event observable in the following way: for each event,
we extract the smallest value of $\tilde{m}_{\tau \tau}$ that gives a
local maximum of the event likelihood and define this to be the
event value of the observable $m_{\mathrm{SV}}$.\footnote{Since we
  have to solve a quartic equation, and since
  real roots thereof come in pairs, we invariably find multiple local maxima
  in the event likelihood.} Our simulations suggest that this
observable gives distributions for Higgs and $Z$ boson event samples
 whose peak locations provide 
 a good
 determination of the
 corresponding boson mass, with small tails. In any case,
 the presence of such effects can be mitigated by comparing
 experimental distributions of
 observables to template Monte-Carlo samples, as we do in simulations
 of pseudo-experiments below.

\section{Simulations and results \label{sec:results}}
Our study is based on a sample from the
{\tt Herwig++} event generator~\cite{Bahr:2008pv},
version 2.52~\cite{Gieseke:2011na} and corresponds to an integrated
luminosity of 20 fb$^{-1}$, 
which is very similar to what has been achieved at the LHC
by the end of 2012.

As regards the signal, recent Standard Model predictions for Higgs production at the LHC may
be found in Ref.~\cite{Dittmaier:2011ti}.   For a Higgs mass of 125
GeV, at a collision energy of 8 TeV, the expected total
cross section is 19.52 pb in the gluon fusion channel and 1.58 pb in
the vector boson fusion channel, with a probable uncertainty of around 10\%.
The predicted SM branching ratio for $\tau\tau$ decay, also given in
Ref.~\cite{Dittmaier:2011ti}, is 6.4\%.
Higgs production followed by $\tau\tau$
decay thus
corresponds to a cross section of 1.34 pb.

For the $Z$ background, CMS \cite{CMS10} reports a flavour-averaged
prediction of $\sigma (pp \rightarrow Z \rightarrow ll) = 1.13$ nb and
a measurement of 1.12 nb for the 8 TeV LHC. We take $\sigma
(pp \rightarrow Z \rightarrow \tau \tau ) = 1.13$ nb. For the $W+j$
background, we use  $\sigma (pp \rightarrow W j) = 2.15$ nb, taken from
{\tt Herwig++}.

All of our simulations are carried out at the parton level, without
showering or hadronization effects,
apart from hadronic tau decays which we simulate using 
{\tt Herwig++}~\cite{Grellscheid:2007tt}. 
The detector response was modelled
as follows.

Firstly we assume identification efficiencies of 0.4 and 0.3 for the 1- and 3-prong hadronic taus, respectively \cite{Aad:2011kt}.
The lepton identification efficiency is assumed to be 0.9 in our analysis.
To estimate the number of $W+j$ events in which the jet mimics a 1- or 3-prong hadronic tau, we use the fake rates of 0.01 and 0.002 for 1- and 3-prong taus, respectively.
These numbers are obtained from the simulation of $W + j$ events \cite{Chatrchyan:2012zz}.

Secondly, we parameterize detector mismeasurements by smearing the energy component of jets and leptons 
with $\sigma(E)/E_j = 0.5\,{\rm GeV}^{\frac{1}{2}} / \sqrt{E_j}$ and  $\sigma(E)/E_\ell = 0.05\,{\rm GeV}^{\frac{1}{2}} / \sqrt{E_\ell}$,
respectively.
For the missing transverse momentum, we smear each component with $\sigma_x = \sigma_y = 5$\,GeV.

For the $\tau$ decay vertex, the Monte-Carlo truth position is smeared
by Gaussian distributions of widths $0.613 \pm 0.008 $ mm and 
$10.5 \pm 0.2 \,\mu$m, in the directions parallel and perpendicular to the
3-prong tau-jet, respectively. For jets that fake taus, we take
the truth vertex position to be zero and then smear as above.

We then apply event selection cuts to purify the signal, for which we impose
\beq
p_{Tj} > 20\,{\rm GeV},~~~~
p_{Tj'} > 20\,{\rm GeV},~~~~
\slashed{p}_T > 20\,{\rm GeV}.
\eeq
For the hadron-lepton mode, we further impose
\beq
m_T \equiv \sqrt{ 2 |{\bf p}^{\ell}_T| |{\vpm_T}| (1 - \cos\Delta \phi)  } < 40\,{\rm GeV}
\eeq
to reduce the background involving $W$s, where $\Delta \phi$ is the azimuthal difference between the lepton and the missing transverse momentum.    
The cross section for each process/channel after taking account of the efficiencies of (mis)identification and the selection cuts
is listed in Table \ref{tab:xsec}.
We use the 3 prong-hadron\,(-lepton) channel for $m_{\mathrm SV}$ and 
the hadron-hadron\,(-lepton) channel for $m_{\mathrm vis}$ and $m_{\mathrm eff}$. 

\begin{table}[t!]
\begin{center}
\begin{tabular}{|c||c|c||c|c|}
\hline
& 3pr-had & 3pr-lep & had-had & had-lep \\  \hline
$H$ & 6.3 & 5.7 & 17.2 & 28.5 \\ \hline
$Z$ & 528.8 & 416.5 & 1451.4 & 2036.0 \\ \hline
$Wj$ & 50.5 & 32.4 & 135.3 & 161.9 \\ \hline
\end{tabular} 
\caption{ The cross section times efficiency of each process/channel in fb.}
\label{tab:xsec}
\end{center}
\end{table}
%

\begin{figure}
\begin{minipage}{0.5\linewidth}
\centering
\includegraphics[width=0.9\textwidth]{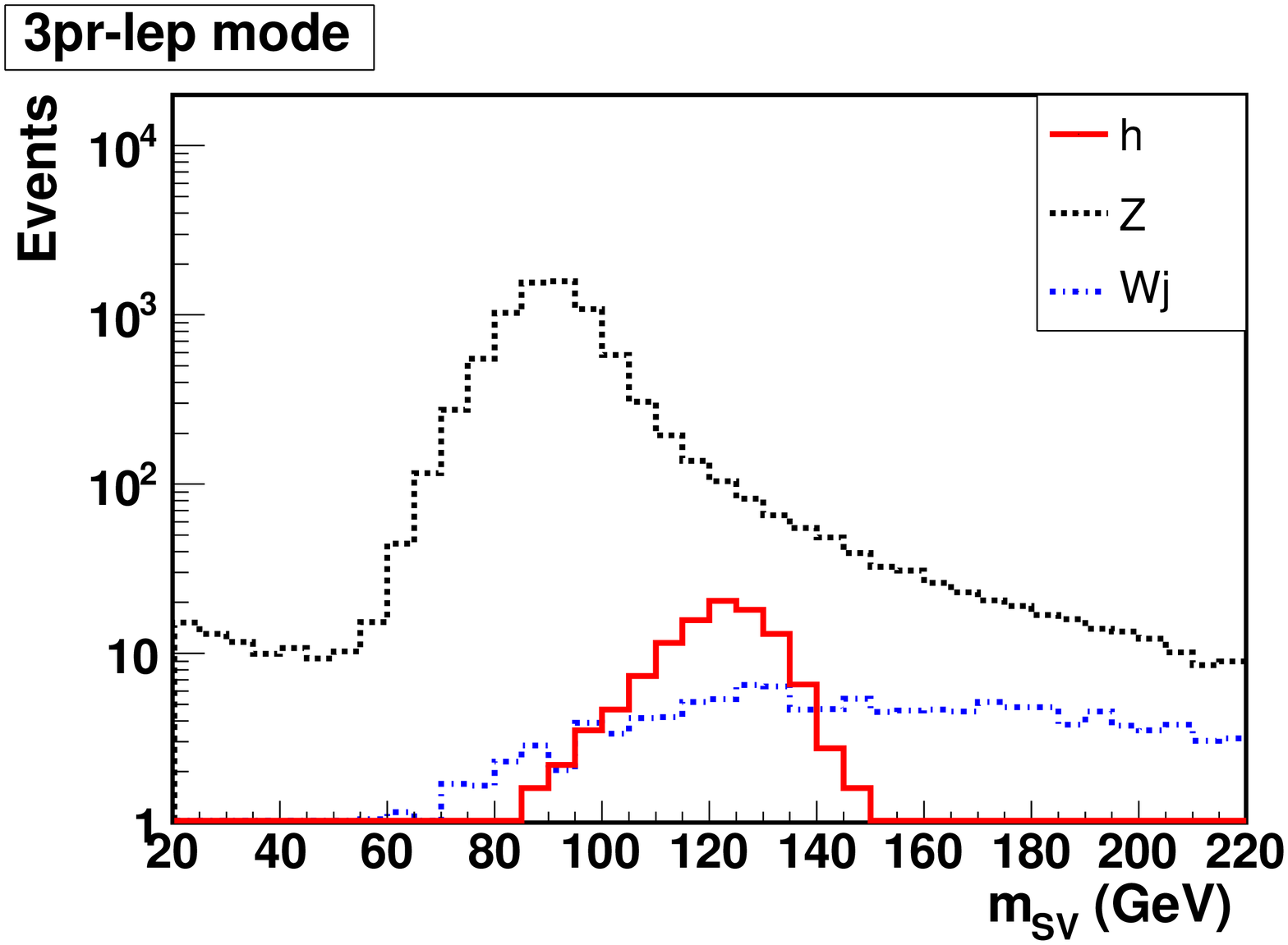}
\end{minipage}
\begin{minipage}{0.5\linewidth}
\centering
\includegraphics[ width=0.9\textwidth]{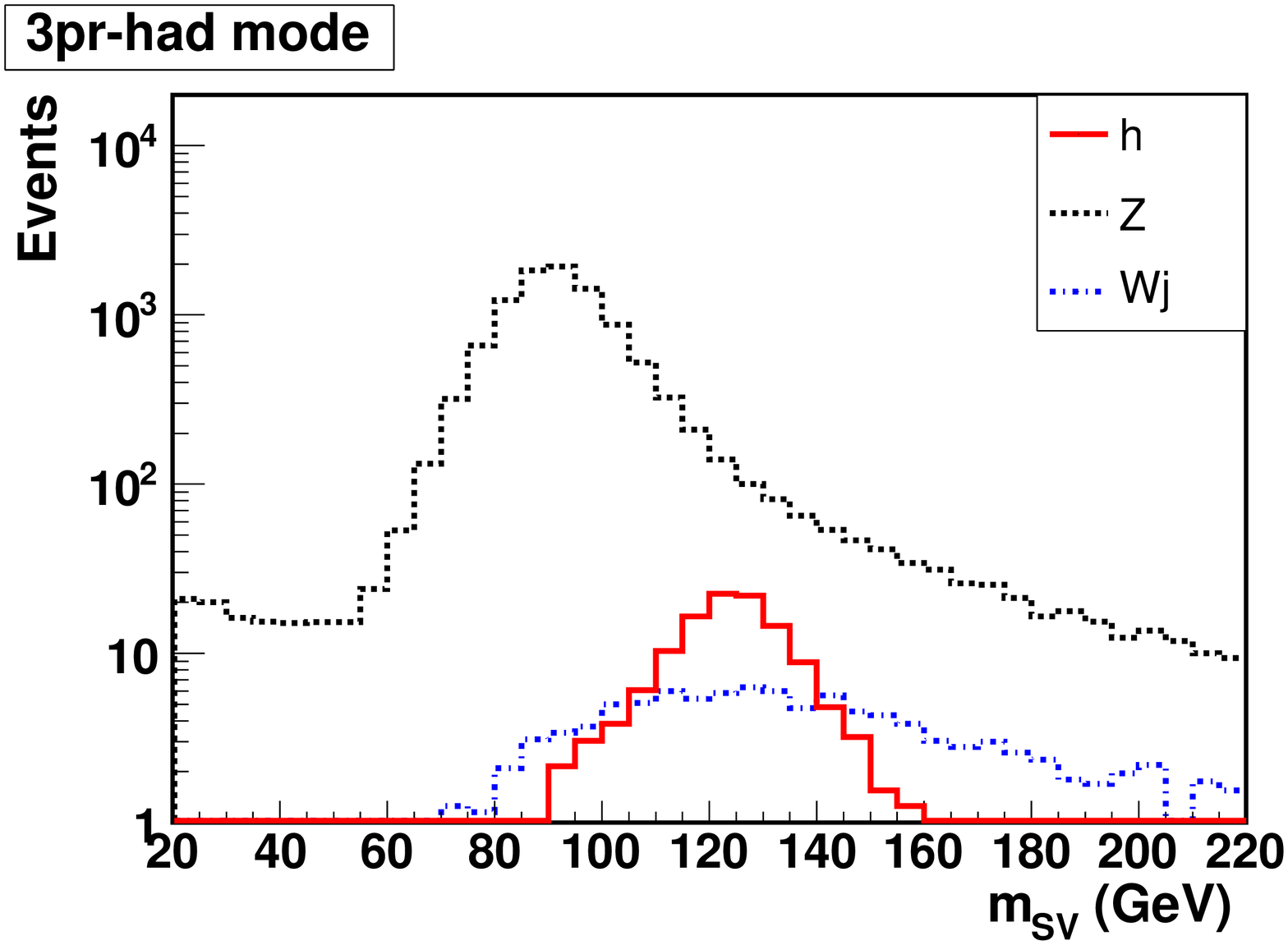}
\end{minipage}

\begin{minipage}{0.5\linewidth}
\centering
\includegraphics[ width=0.9\textwidth]{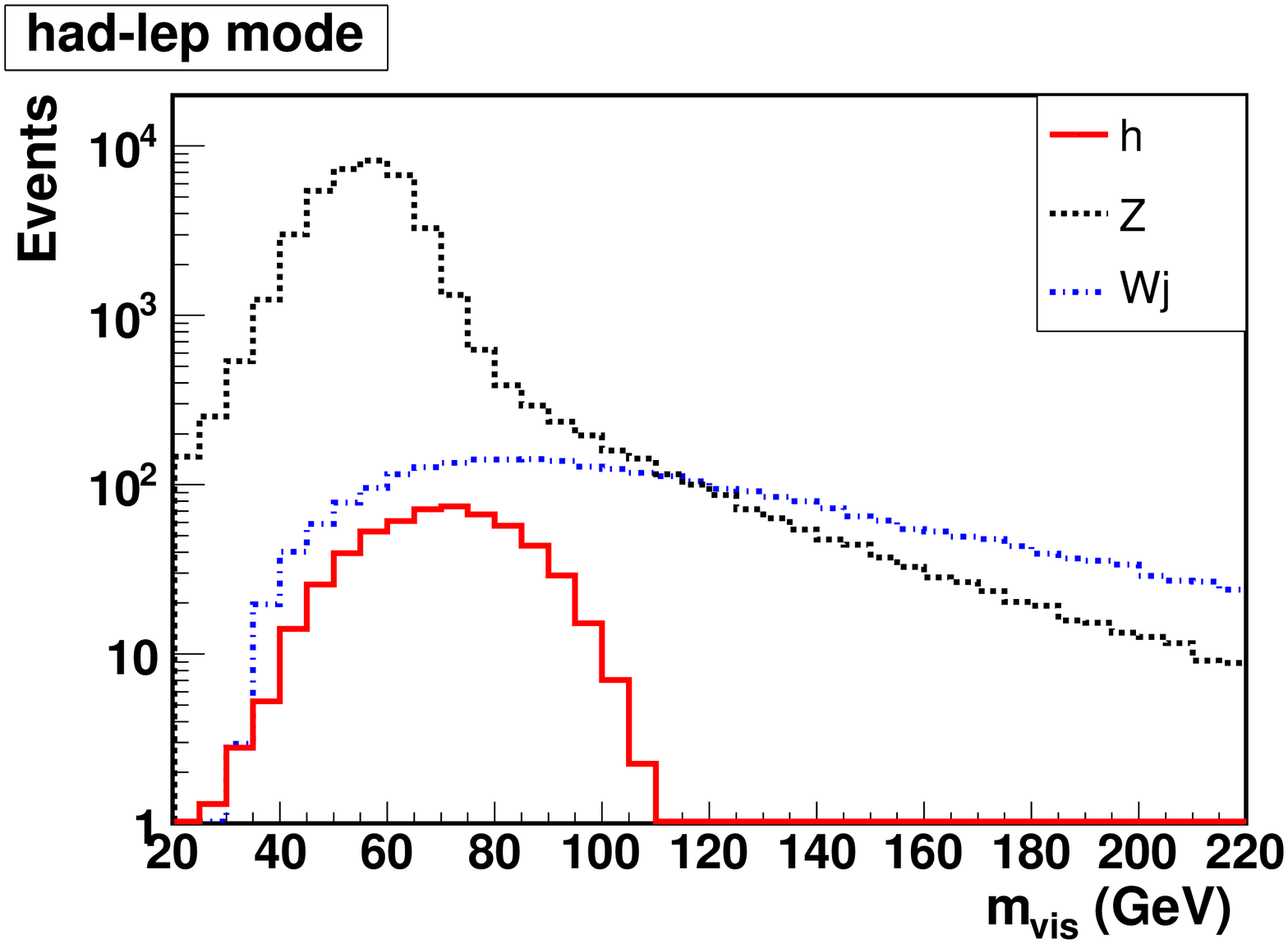}
\end{minipage}
\begin{minipage}{0.5\linewidth}
\centering
\includegraphics[ width=0.9\textwidth]{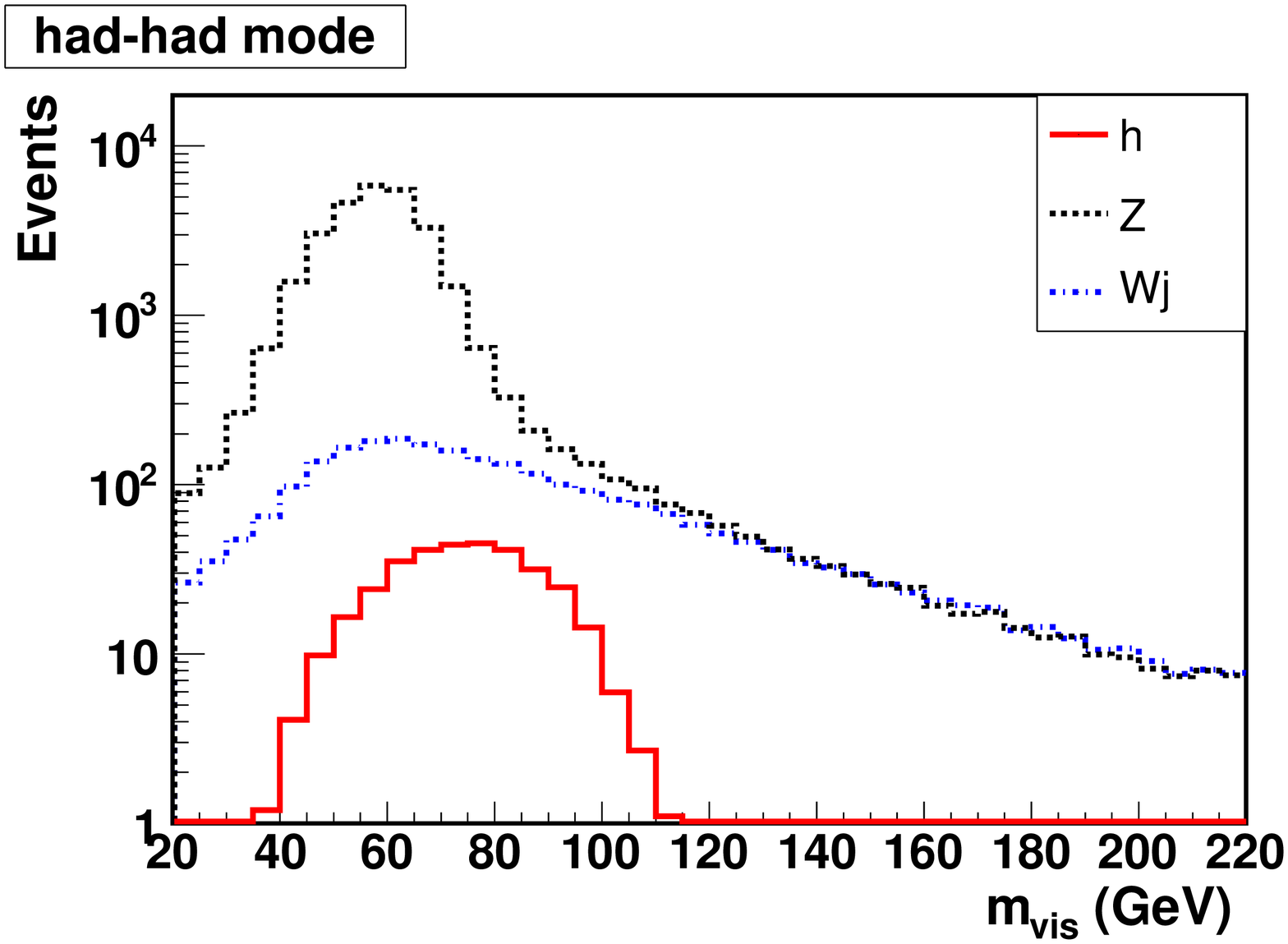}
\end{minipage}

\begin{minipage}{0.5\linewidth}
\centering
\includegraphics[ width=0.9\textwidth]{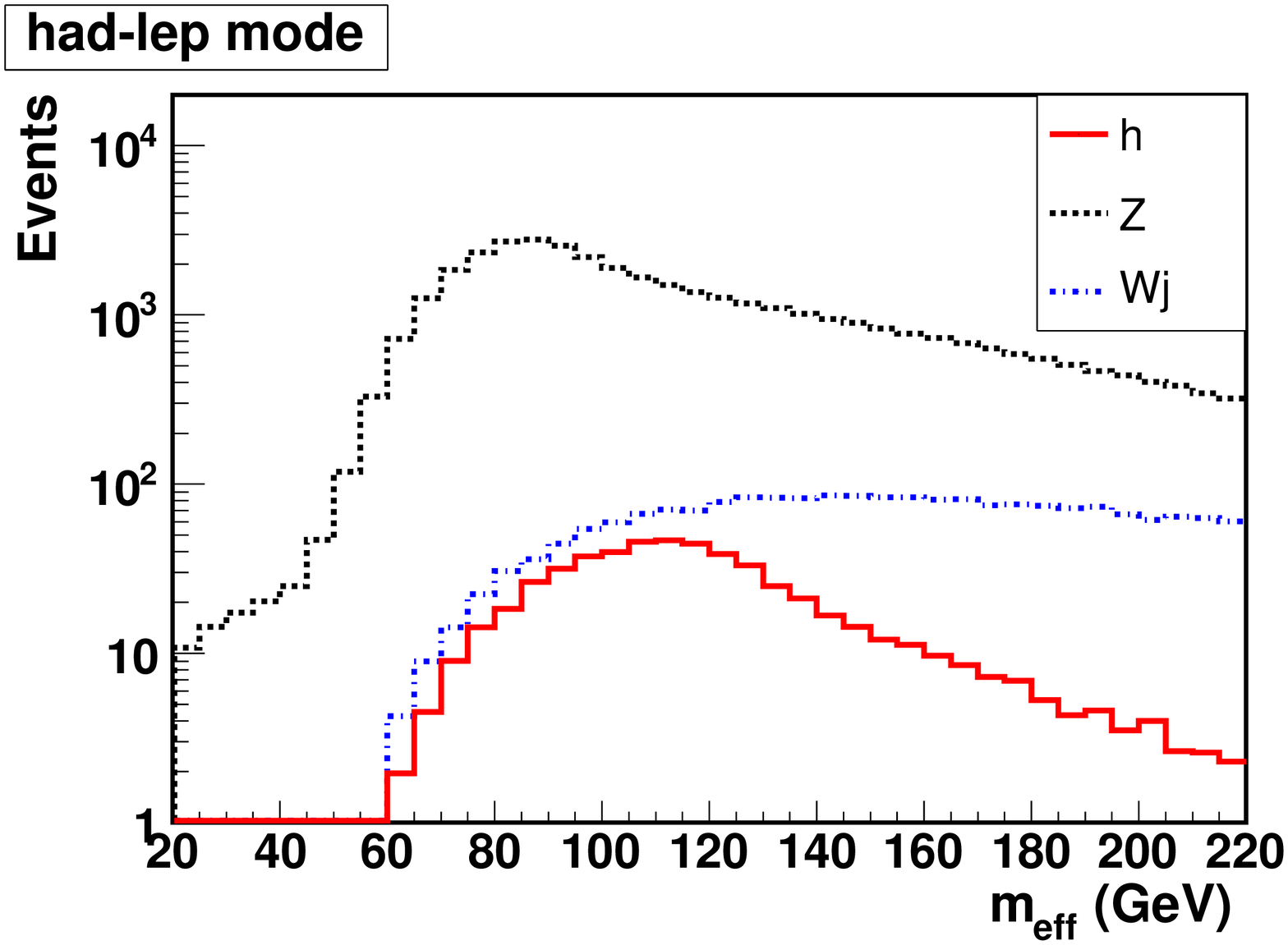}
\end{minipage}
\begin{minipage}{0.5\linewidth}
\centering
\includegraphics[ width=0.9\textwidth]{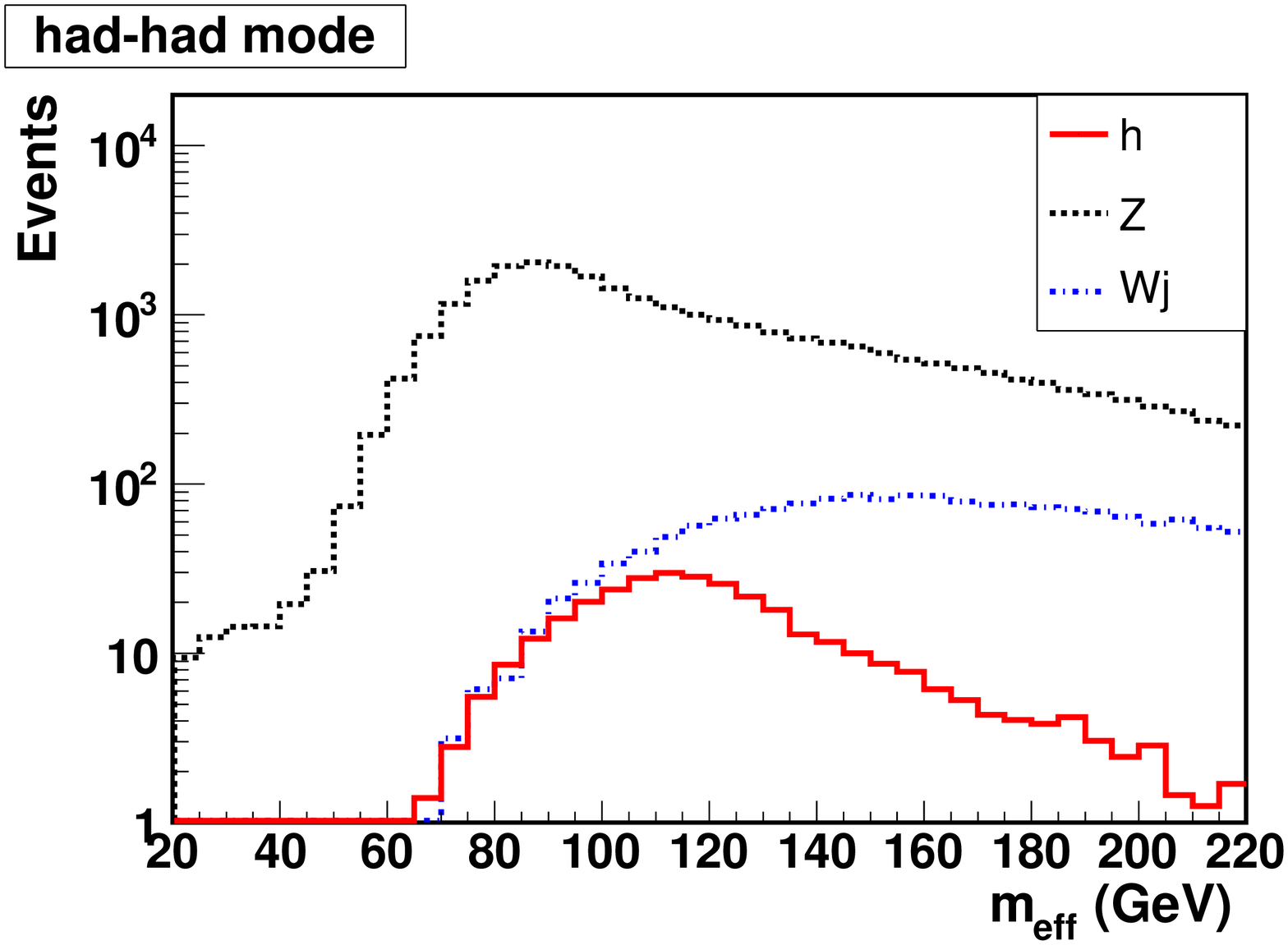}
\end{minipage}
\caption{Signal and background distributions of $m_{\mathrm{SV}}$
  (top), $m_{\mathrm{vis}}$ (centre) and  $m_{\mathrm{eff}}$ (bottom), for an integrated
  luminosity of 20 fb$^{-1}$ at the 8 TeV LHC, for $m_h =125$
  GeV. Left: lepton-hadron modes; right: hadron-hadron modes. \label{fig:distros} }
\end{figure}

In Figure \ref{fig:distros} we compare the signal and background
distributions of our variable $m_{\mathrm{SV}}$ with the existing
variables $m_{\mathrm{eff}}$ and
$m_{\mathrm{vis}}$. $m_{\mathrm{vis}}$ is simply the invariant mass of
the visible products of both $\tau$ decays, while $m_{\mathrm{eff}}$
includes the missing transverse momentum (an explicit definition may
be found in \cite{Barr:2011he}).
We show results for the lepton-hadron modes and hadron-hadron modes
separately.
The better separation between signal and
backgrounds that we expected to obtain using $m_{\mathrm{SV}}$ is
clear to see in the Figure. Indeed, $m_{\mathrm{vis}}$ has
distributions for the Higgs signal and $Z$ background which are
strongly peaked, but the two peaks sit on top of each
other. $m_{\mathrm{eff}}$
incorporates extra information in the form of the missing transverse
momentum, and slightly increases the separation between the maxima of
the Higgs and $Z$ boson peaks, but at the cost of introducing large
tails (from the smearing of the missing transverse momentum
measurement). As a result, the Higgs signal is easily hidden in the
large tail of the $Z$ background. Moreover, the shapes of all
components become similar, making discrimination difficult when the
overall normalizations are uncertain. 
In contrast $m_\mathrm{SV}$ provides
a good separation between the narrow Higgs and $Z$ boson peaks, which appear at the true mass values.
These
peaks are, furthermore, very different in shape from the continuum $W+$
jet background.

To see how well each variable can reconstruct the resonance, we list 
the peak location of the each variable's distribution and the input mass of the resonance in
Tables~\ref{tab:bias_had} and \ref{tab:bias_lep} in the hadron-hadron and hadron-lepton modes, respectively.
The pure $Z \to \tau \tau$ and $h \to \tau \tau$ samples with several Higgs masses are used.
The peak location is calculated as the weighed average of the three highest bins.
The error is estimated by using 10 independent samples.
The tables show clearly that $m_\mathrm{SV}$ reconstructs the masses of resonances very well compared to the other variables.  Although this correlation is not the basis of our method  for mass determination, it helps to separate the signal from the background.

Another remark is that the $S/B$ for the $W+$ jet background is better for $m_\mathrm{SV}$, 
as can be seen from Fig.~\ref{fig:distros}.
This is because a fraction of the $W+$ jet background events do not produce real solutions.
The fact that the tau mass is much smaller than the typical momentum scale of the reconstructed objects 
(jets and leptons) implies that the neutrino momenta are inferred to be very close 
to those objects (see e.g. Eq.~(\ref{eq:reconptau})).
However the direction of this inferred momentum tends to conflict with the direction of the observed missing transverse momentum, since the neutrino from the $W$ decay is generally not collimated with respect to those objects, leading to no real solution. 

Nevertheless, it is also apparent that the statistics available using
$m_\mathrm{SV}$ are lower than for the other variables, even in the region of maximum signal.
We need, therefore, to make a quantitative comparison of the three variables. To
do so,
we generate distributions of them (for $m_h = 125$ GeV signal and backgrounds) in
ten pseudo-experiments, each corresponding to an integrated
luminosity of 20 fb$^{-1}$ of 8 TeV LHC
data. Each pseudo-experiment is then compared to template model distributions
with different values of $m_h$ and with different normalization
factors $f_h$, $f_W$, and $f_Z$ for the Higgs signal and $W+j$ and $Z$
backgrounds, respectively. 
(The values  $f_{h,W,Z} =1$ correspond to the leading order Monte-Carlo prediction.) 
Allowing the model distribution normalizations to float in this way allows us not only to take into
account some of the most important systematic effects\footnote{
We assume uniform probability distribution functions for $f_{h,W,Z}$ in the likelihood calculation.
In the actual experimental situation, these probabilities are not uniform but localised around $f_{h,W,Z} = 1$
to avoid too large/small values.
In this sense, our treatment of the systematic uncertainty is conservative.   
} 
(arising from the uncertainties in the luminosity, the Monte-Carlo predictions, and data-driven extrapolations), but also provides a means to measure the
cross-section times branching ratio for Higgs production followed by
decay to $\tau \tau$.


\begin{table}
\begin{center}
\begin{tabular}{|c|c|c|c|c|}
\hline
& $Z$(91) & $h$(119) & $h$(125) & $h$(131) \\ \hline
$m_{\rm SV}$ & $92.10 \pm 0.01$ & $117.81 \pm 0.73$ & $124.75 \pm 0.75$ & $130.84 \pm 1.00$  \\ \hline
$m_{\rm eff}$  & $89.97 \pm 0.02$ & $110.55 \pm 1.09$ & $115.12 \pm 1.19$ & $115.12 \pm 1.19$ \\ \hline
$m_{\rm vis}$  & $60.27 \pm 0.01$ & $73.87 \pm 0.84$ & $77.82 \pm 1.05$ & $78.42 +- 0.98$  \\ \hline
\end{tabular}
\caption{ Peak position vs.~input mass in the hadron-hadron mode.
The numbers in the parentheses are the input mass of the bosons.  
Pure background and signal samples are used.  
\label{tab:bias_had} }
\end{center}
%
%
\begin{center}
\begin{tabular}{|c|c|c|c|c|}
\hline
& $Z$(91) & $h$(119) & $h$(125) & $h$(131) \\ \hline
$m_{\rm SV}$ & $91.54 \pm 0.40$ & $116.12 \pm 0.82$ & $122.15 \pm 0.84$ & $129.33 \pm 0.68$  \\ \hline
$m_{\rm eff}$  & $89.90 \pm 0.02$ & $108.64 \pm 1.12$ & $113.27 \pm 0.96$ & $116.38 \pm 1.20$ \\ \hline
$m_{\rm vis}$  & $59.87 \pm 0.01$ & $70.82 \pm 0.83$ & $73.59 \pm 0.77$ & $75.52 \pm 1.28$  \\ \hline
\end{tabular}
\caption{
Peak position vs.~input mass in the hadron-lepton mode.
The numbers in the parentheses are the input mass of the bosons.  
Pure background and signal samples are used.  
\label{tab:bias_lep} }
\end{center}
\label{default}
\end{table}%


We make a cut on the observable of interest itself, so as
to maximize its discovery potential. Roughly
speaking, this cut selects the region that contains the bulk of
the signal events for that observable. Thus we choose
$m_\mathrm{SV} \in  [110, 150]$ GeV,
$m_\mathrm{vis} \in  [75, 110]$ GeV, and
$m_\mathrm{eff} \in  [100, 150]$ GeV.

We then compute, for
each pseudo-experiment, a binned Poisson log-likelihood, $\log \mathcal{L}
= \Sigma_i \log
\mathcal{L}^P\left( n^\mathrm{trial}_i; x_i \right)$,
where $n^\mathrm{trial}_i$ is the number of events observed in
histogram bin $i$ in a given pseudo-experiment and $x_i = x_i (m_h,f_h, f_Z, f_W)$ is the number of events expected in a
given model, parameterized by $m_h,f_h, f_Z, f_W$.

\begin{figure}
\begin{minipage}[b]{0.5\linewidth}
\centering
\includegraphics[width=0.9\textwidth]{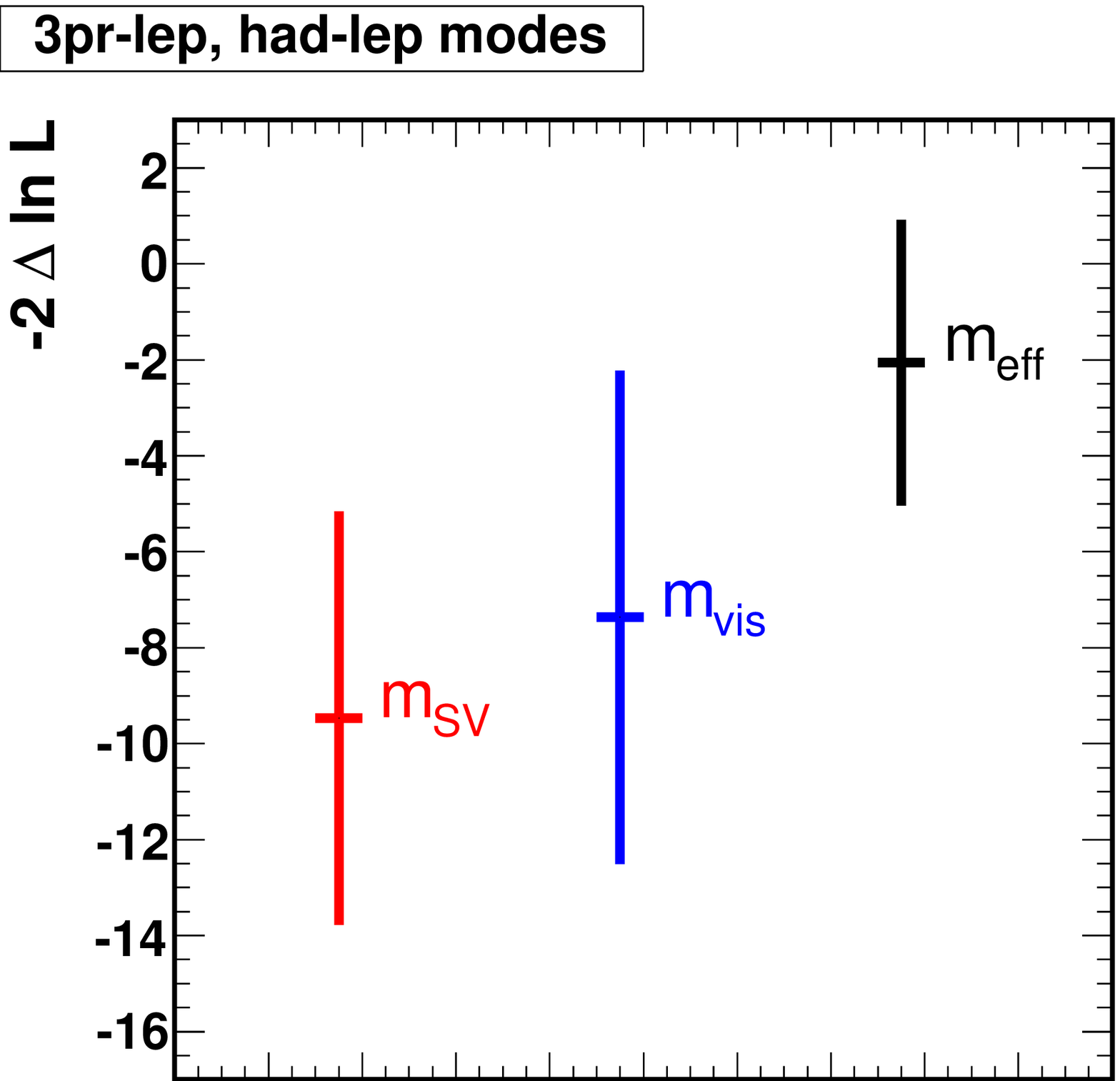}
\end{minipage}
\begin{minipage}[b]{0.5\linewidth}
\centering
\includegraphics[width=0.9\textwidth]{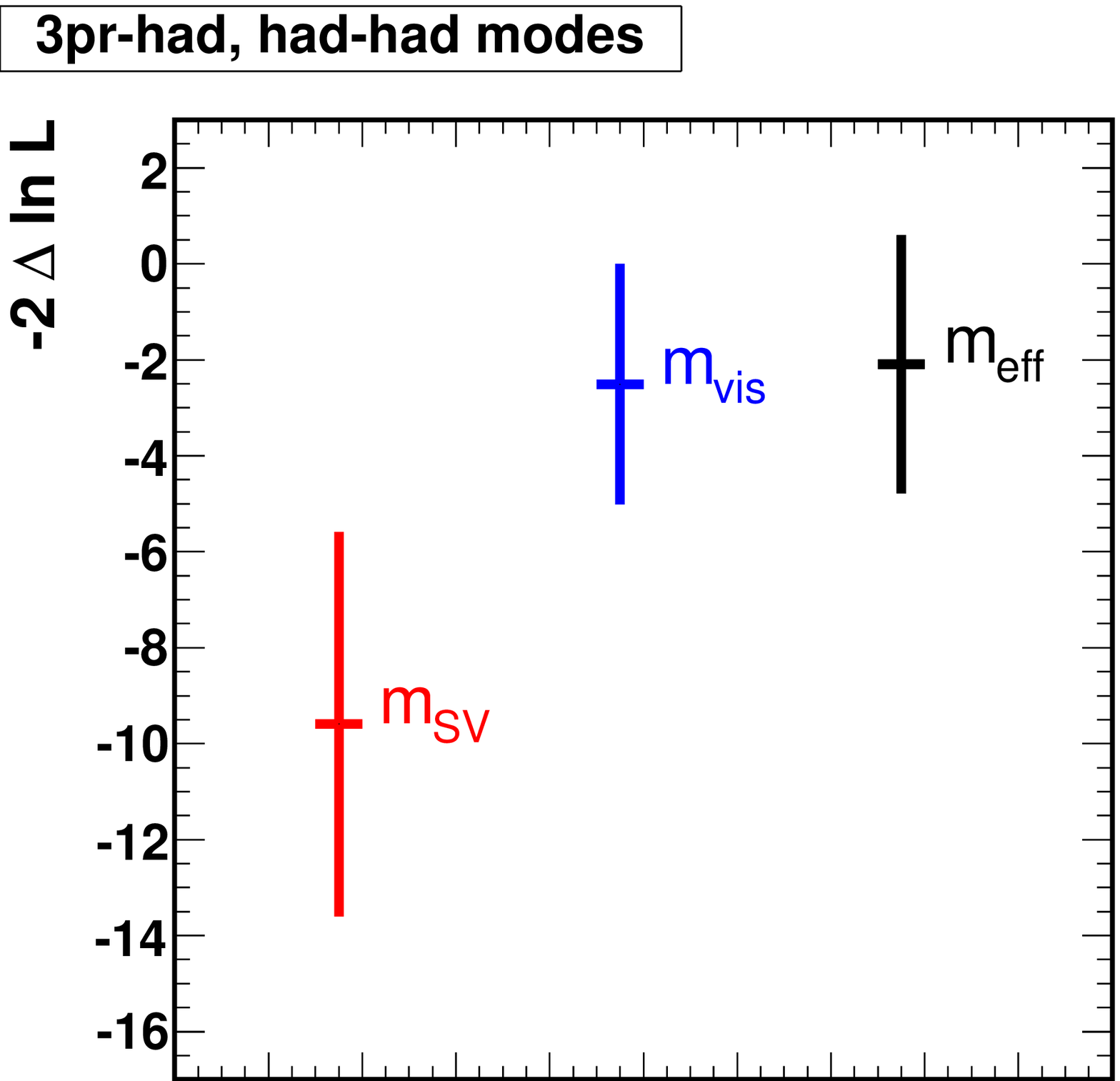}
\end{minipage}
\caption{Discovery potential using $m_{\mathrm{SV}}$, compared to
  $m_{\mathrm{eff}}$ and  $m_{\mathrm{vis}}$, for lepton-hadron (left)
  and hadron-hadron (right) modes, for an integrated
  luminosity of 20 fb$^{-1}$ at the 8 TeV LHC. The centre of each bar shows the mean value over trials,
while the width of each bar gives the root-mean-square deviation. \label{fig:discovery} }
\end{figure}

To assess the discovery potential, we then compute the difference in
log-likelihood between models with and without a Higgs signal. In the
model without a signal, we maximize the log-likelihood with respect to
$f_Z,$ and $f_W$, whereas in the model with a signal, we additionally
maximize with respect to $m_h$ and $f_h$. In Fig.~\ref{fig:discovery}, 
we show $-2$ times the difference in log-likelihood for the three
variables. The centre of each bar shows the mean value over trials,
while the width of each bar gives the root-mean-square deviation over
trials. 
\begin{figure}
\begin{minipage}[b]{0.5\linewidth}
\centering
\includegraphics[width=0.9\textwidth]{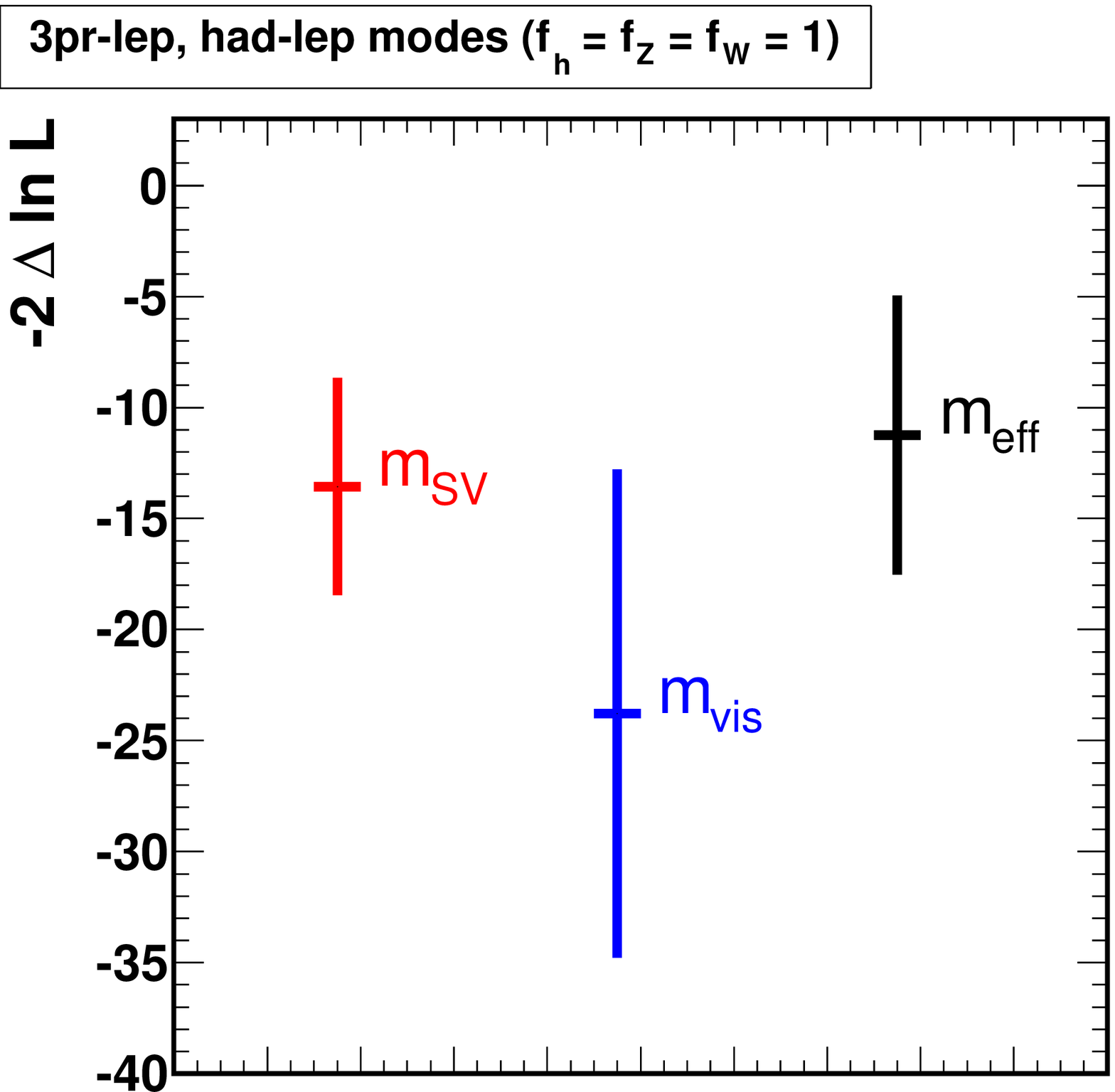}
\end{minipage}
\begin{minipage}[b]{0.5\linewidth}
\centering
\includegraphics[width=0.9\textwidth]{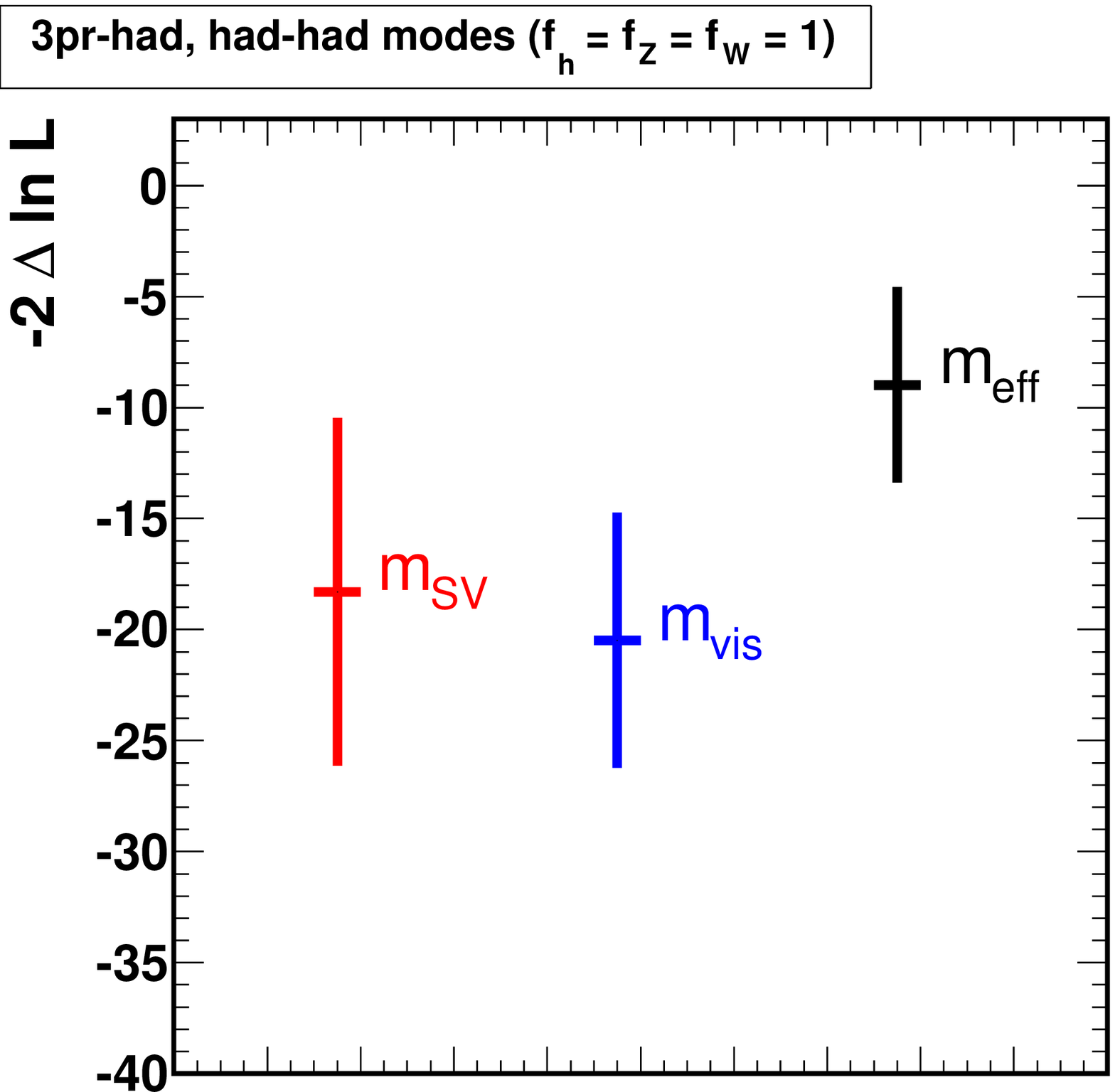}
\end{minipage}
\caption{
  The same as Fig.~2, however $f$s are fixed at 1 in the likelihood calculation.
  }
  \label{fig:discovery_fix}
\end{figure}
\begin{figure}
\begin{minipage}[b]{0.5\linewidth}
\centering
\includegraphics[width=0.9\textwidth]{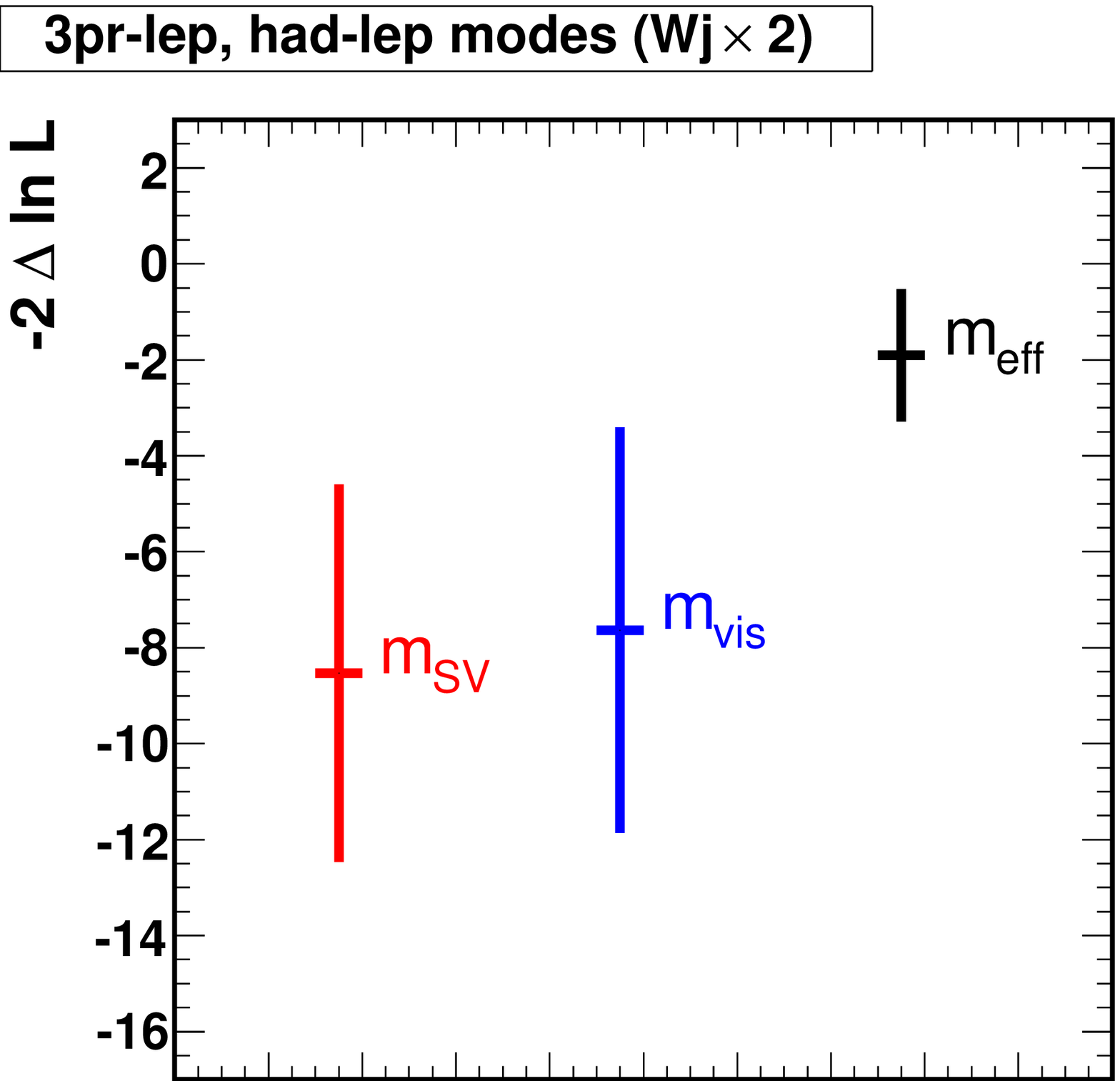}
\end{minipage}
\begin{minipage}[b]{0.5\linewidth}
\centering
\includegraphics[width=0.9\textwidth]{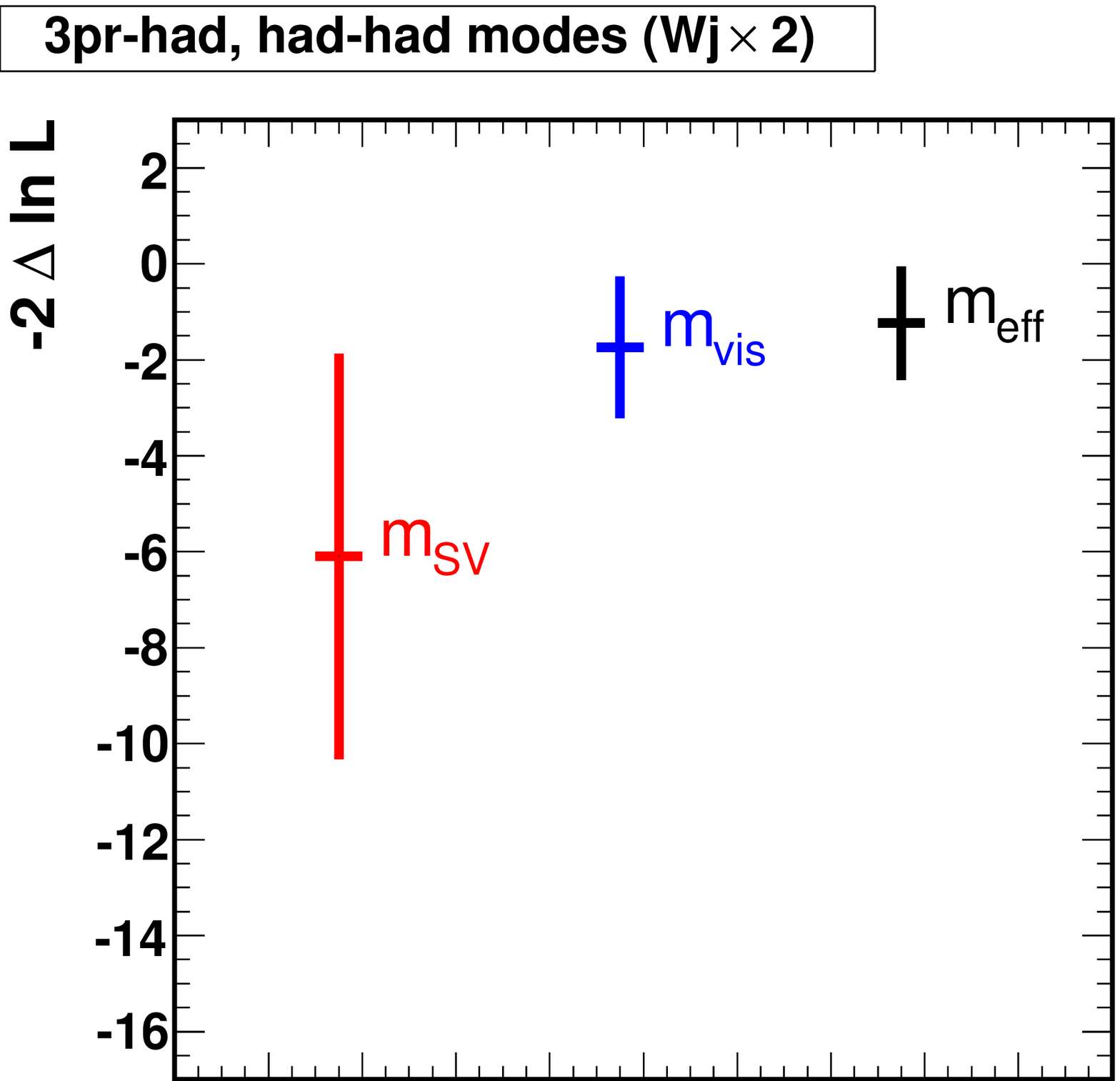}
\end{minipage}
\caption{
	The same as Fig.~2, however 
  the analysis uses a $W+j$ background sample which is twice as large as the expected one. 
  }
  \label{fig:discovery_w2}
\end{figure}

We observe a significant improvement using $m_\mathrm{SV}$ in the
hadron-hadron channel, along with a more modest improvement (compared
to $m_\mathrm{vis}$) in the leptonic channel. 
The different performances of $m_\mathrm{vis}$ in hadron-lepton and hadron-hadron modes can be understood from the distributions in Fig.~\ref{fig:distros}. 
Unlike the hadron-lepton mode, the $m_\mathrm{vis}$ distribution in the hadron-hadron mode shows that both $Z$ and $W+j$ backgrounds as well as the signal have falling shapes in the signal region $m_\mathrm{vis} \in  [75, 110]$\,GeV.
This suggests that in the hadron-hadron mode, fitting the signal + background distribution with only the $Z$ and $W+j$ backgrounds by floating $f_Z$ and $f_W$ can be more possible compared to the hadron-lepton mode. 
This feature can be seen explicitly in Fig.~\ref{fig:discovery_fix}, which show the same likelihoods as in Fig.~\ref{fig:discovery}
but with the $f$s fixed at 1. 
Floating the $f$s brings significant degradation for $m_\mathrm{vis}$ in the hadron-hadron mode.

The absolute values of the discovery significance are
exaggerated, since we have neglected sub-dominant backgrounds and many
uncertainties, but the relative performance of the different variables
should be meaningful.

We have estimated the size of $W + j$ background using the reported tau fake rates in $W + j$ events.
The tau fake rate is generally dependent on the tau identification algorithm and the jet $p_T$.
To check the robustness of our result, in Fig.~\ref{fig:discovery_w2}
we show the same discovery potential plots as Fig.~\ref{fig:discovery}
but containing a $W + j$ background twice as large as the one in Fig.~\ref{fig:discovery}.
The discovery potentials are degraded slightly but the qualitative features are unchanged.

\begin{figure}
\begin{minipage}[b]{0.5\linewidth}
\centering
\includegraphics[ width=0.9\textwidth]{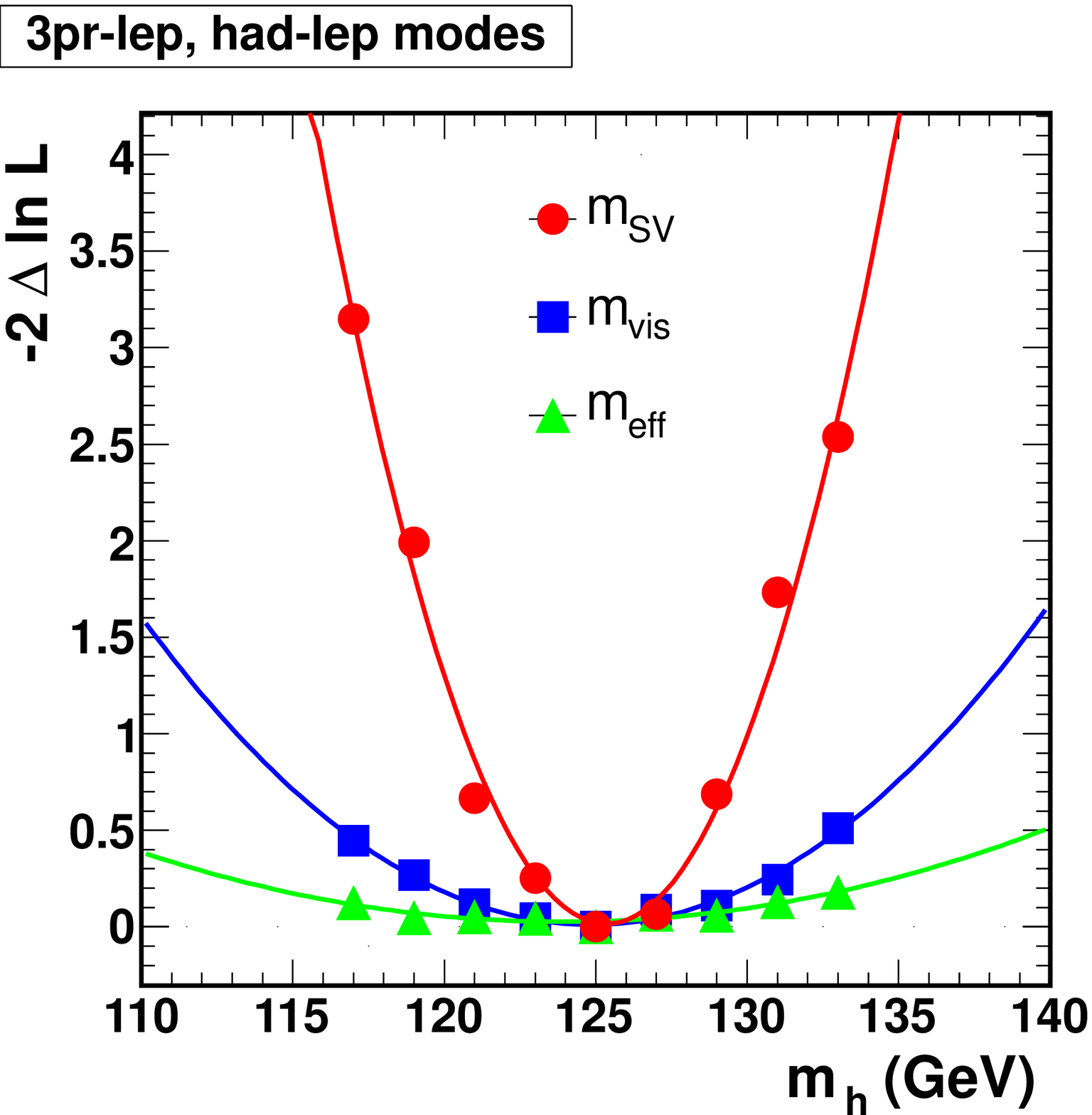}
\end{minipage}
\begin{minipage}[b]{0.5\linewidth}
\centering
\includegraphics[ width=0.9\textwidth]{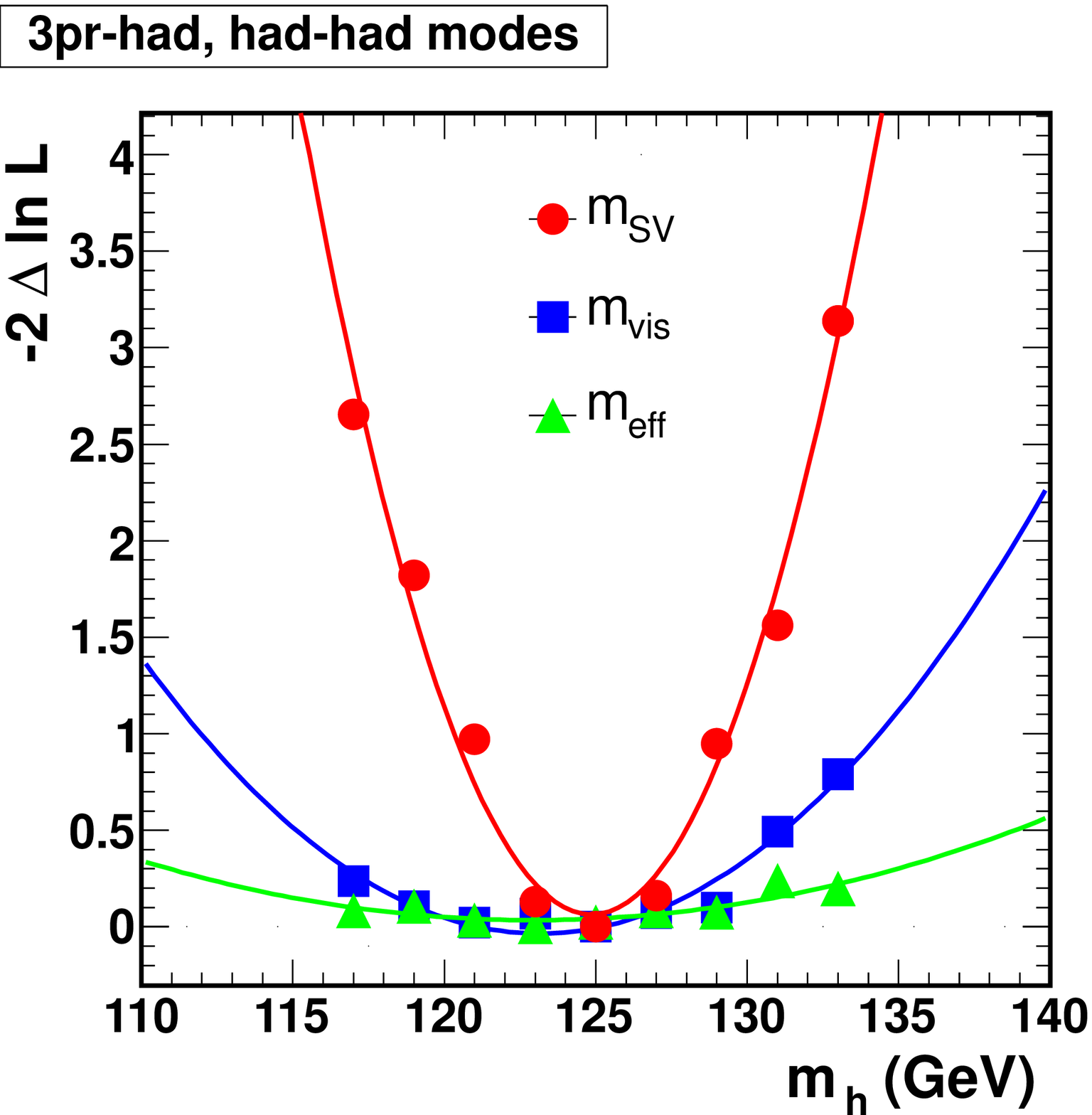}
\end{minipage}
\caption{The log-likelihood function near $m_h = 125$ GeV obtained
  using the $m_{\mathrm{SV}}$,
  $m_{\mathrm{eff}}$ and  $m_{\mathrm{vis}}$ distributions, for lepton-hadron (left)
  and hadron-hadron (right) modes, for an integrated
  luminosity of 20 fb$^{-1}$ at the 8 TeV LHC.\label{fig:mass1} }
\end{figure}
\begin{figure}
\begin{minipage}[b]{0.5\linewidth}
\centering
\includegraphics[ width=0.9\textwidth]{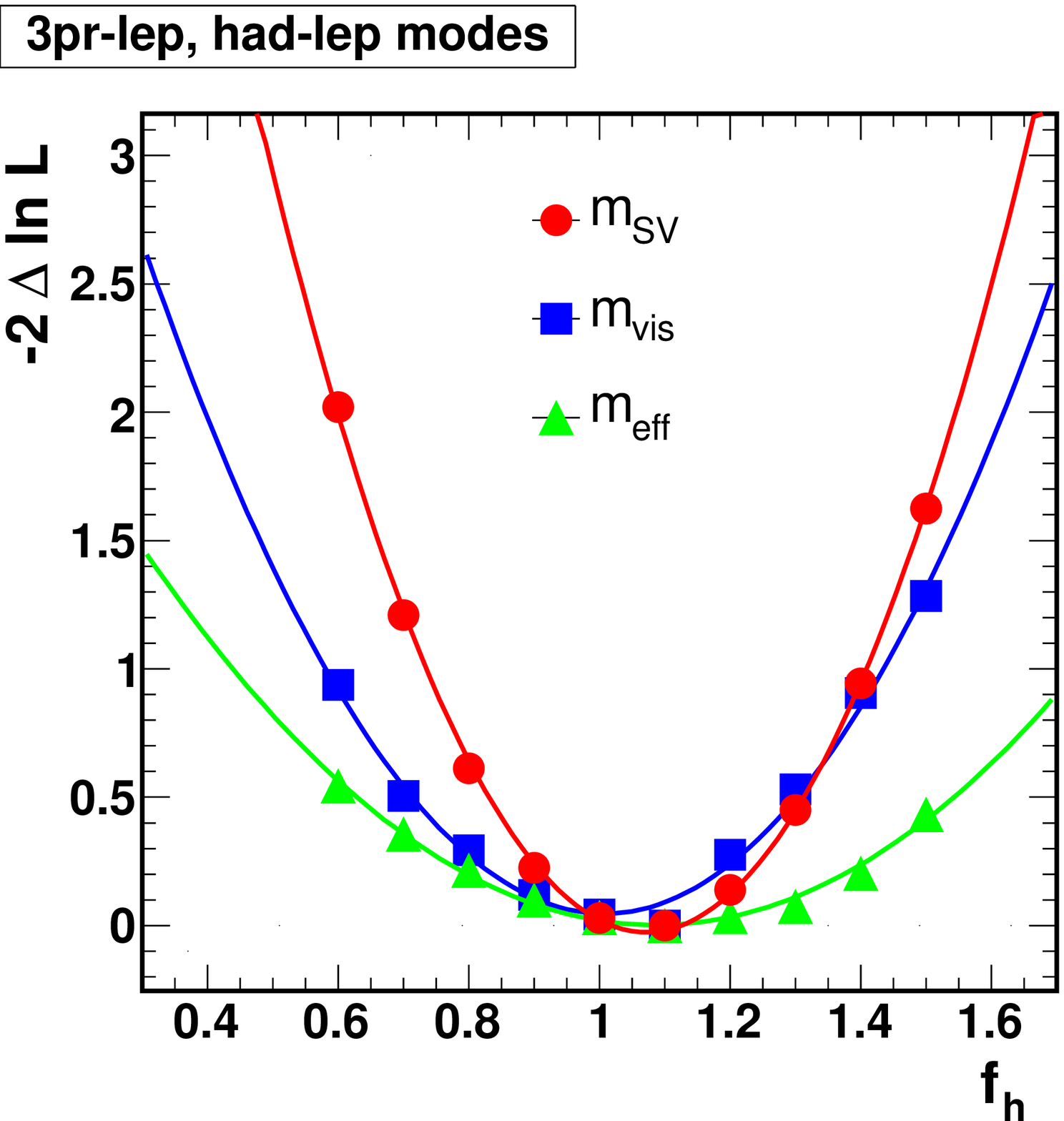}
\end{minipage}
\begin{minipage}[b]{0.5\linewidth}
\centering
\includegraphics[width=0.9\textwidth]{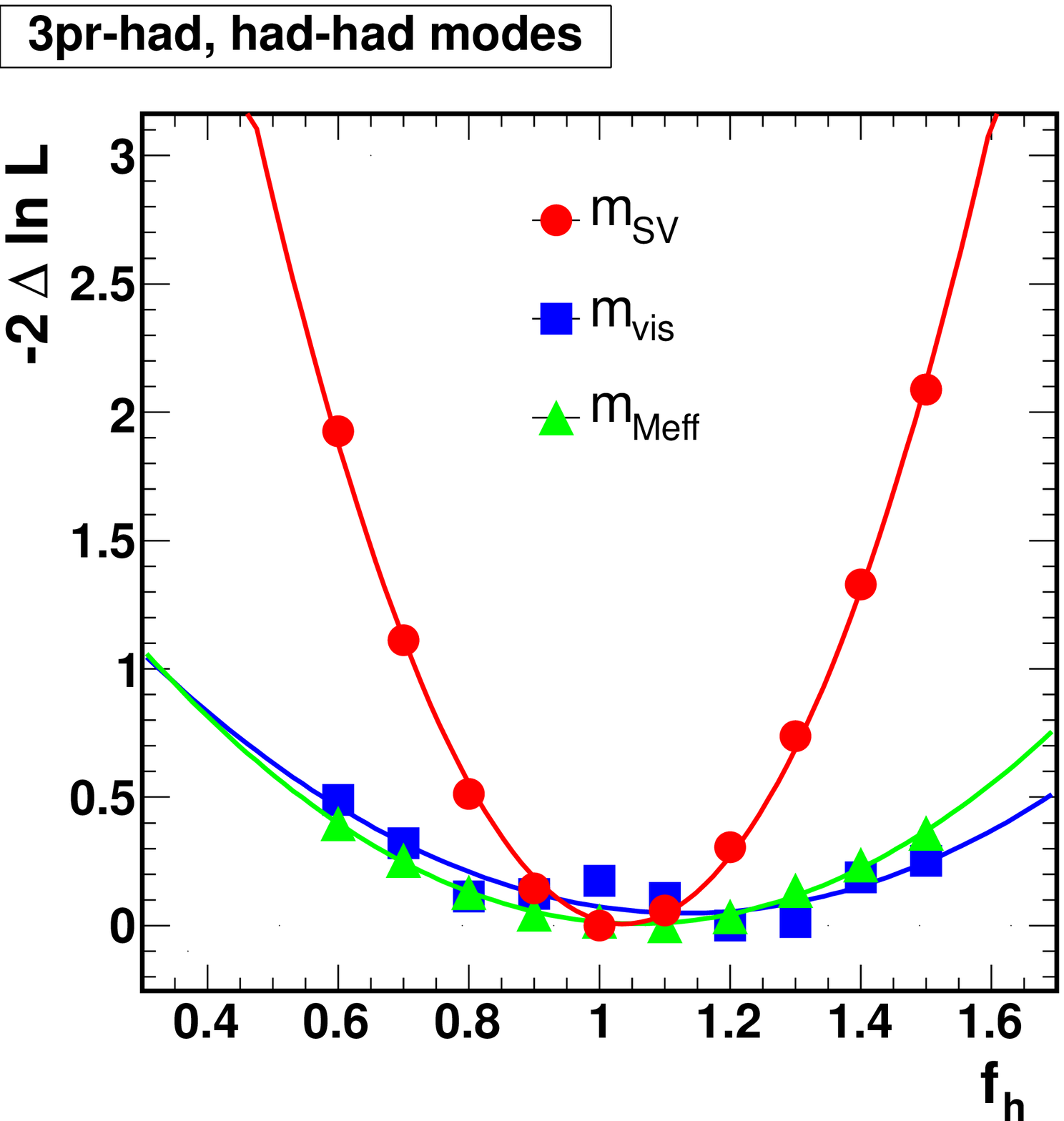}
\end{minipage}
\caption{The log-likelihood as a function of the signal strength $f_h$, obtained
  using the $m_{\mathrm{SV}}$,
  $m_{\mathrm{eff}}$ and  $m_{\mathrm{vis}}$ distributions, for lepton-hadron (left)
  and hadron-hadron (right) modes, for an integrated
  luminosity of 20 fb$^{-1}$ at the 8 TeV LHC. $f_h=1$ corresponds to
  the truth value. \label{fig:sigmabr1} }
\end{figure}

\begin{table}
\begin{center}
\begin{tabular}{c|c|c|c}
& $m_{\rm SV}$ & $m_{\rm vis}$ & $m_{\rm eff}$ \\ \hline
had-lep & 4.7 & 11.0 & 23.3 \\ \hline
had-had & 4.7 & 11.8 & 23.1  \\
\end{tabular}
\caption{Uncertainty in the Higgs boson mass (in GeV), as measured
  $m_{\rm SV}$, $m_{\mathrm{eff}}$ and  $m_{\mathrm{vis}}$ distributions, for lepton-hadron
  and hadron-hadron modes, for an integrated
  luminosity of 20 fb$^{-1}$ at the 8 TeV LHC.\label{tab:mass2} }
\end{center}
\label{default}
\end{table}%
To assess the expected resolution in the Higgs mass measurement, we first show, in Fig.~\ref{fig:mass1}, the variation in
$-2\log\mathcal{L}$, averaged over pseudo-experiments, for values of the model Higgs mass, $m_h$,
in the neighbourhood of 125 GeV, after maximizing $\mathcal{L}$ with respect to
$f_h, f_Z$, and $f_W$. In the figure, we have placed the minimum of
$-2\log\mathcal{L}$
at height zero.  
Because the pseudo-experimental data and templates are prepared in the same way, we cannot estimate any biases that might occur when each variable is used to determine the mass from real data. However, for each variable, we can estimate the precision of the mass measurement from a quadratic fit to the log-likelihood.
The resulting fractional uncertainties are shown in Table.~\ref{tab:mass2} and are seen to be much smaller for $m_{\mathrm{SV}}$ than for the other two observables, in both
hadron-hadron and lepton-hadron modes.
This is easily
explained by the fact that the Higgs signal peak is sharpest for $m_{\mathrm{SV}}$.
For the measurement of the product of the production cross section and
the branching ratio for the decay, the procedure is exactly analogous,
except that now we vary $f_h$, after maximizing with respect to
$m_h$. The resulting fit and fractional uncertainties are shown in
Fig.~\ref{fig:sigmabr1} and Table~\ref{tab:sigmabr}, respectively. 
Again, there is always an improvement when using $m_{\mathrm{SV}}$.
For $m_{\mathrm{vis}}$, the difference in resolution between the hadron-lepton and hadron-hadron modes
can be blamed on different shapes in the $W+j$ background in these modes,
as we have discussed earlier in the connection with the different discovery potentials for $m_{\mathrm{vis}}$ 
between these modes.

\begin{table}
\begin{center}
\begin{tabular}{c|c|c|c}
& $m_{\rm SV}$ & $m_{\rm vis}$ & $m_{\rm eff}$ \\ \hline
had-lep & 0.33 & 0.44 & 0.65 \\ \hline
had-had & 0.32 & 0.93 & 0.73  \\
\end{tabular}
\caption{Resolution for measurement of the production cross section times
 branching ratio  
 for $pp \rightarrow h \rightarrow \tau \tau$ normalized by the leading order prediction
 using $m_{\mathrm{SV}}$, compared to
  $m_{\mathrm{eff}}$ and  $m_{\mathrm{vis}}$, for lepton-hadron (left)
  and hadron-hadron (right) modes, for an integrated
  luminosity of 20 fb$^{-1}$ at the 8 TeV LHC.\label{tab:sigmabr}}
\end{center}
\label{default}
\end{table}%

\section{Conclusions  \label{sec:conclusions}}
As shown in Figs.~\ref{fig:discovery}-\ref{fig:sigmabr1} and Tables~\ref{tab:mass2}-\ref{tab:sigmabr}, our
simulations suggest that a significant improvement in discovery
potential, Higgs boson mass resolution, and measurement of production
cross section times branching ratio can be obtained by focussing on
3-prong $\tau$ decays. The performance of $m_{\mathrm{SV}}$  is
roughly comparable irrespective of whether the other $\tau$ meson decays leptonically or
hadronically, leading to greater gains in the hadron-hadron
channel, where the other observables perform more poorly. 
One possible reason for this is that 
the detector resolution is invariably poorer in this channel, in which final-state leptons are replaced by jets.
Thus the performance of $m_{\mathrm{vis}}$ and $m_{\mathrm{eff}}$ is degraded.
However, these events are also
fully reconstructible using the vertex information, in the absence of smearing. As a result,
the likelihood that
defines the observable
$m_{\mathrm{SV}}$ is able to correct for the extra smearing to a certain extent, by insisting that the
unsmeared quantites consistently reconstruct the event.

The gains are greatest for the mass measurement, which is perhaps not
surprising since our method provides a means to reconstruct the mass
whilst partially correcting for the uncertainties that are introduced by
the detector resolution.

The results of our simulations are encouraging, but they should be
taken with a pinch of salt. The simulations themselves are rudimentary, and we have only performed a comparison
with the basic variables $m_{\mathrm{eff}}$ and
$m_{\mathrm{vis}}$. Both collaborations now employ more sophisticated
likelihood-based analyses. Unfortunately the full details of these
have not been made public, so it is difficult for us to make a fair
comparison. 
CMS do say that their likelihood method gives a Higgs mass resolution of around 21$\%$ compared to 24$\%$ using mvis \cite{Chatrchyan:2011nx}.

We have also not made a full study of the
backgrounds, of which many are relevant for this search. However, the two backgrounds we did consider are very different in their
nature (one being a genuine, resonant background and the other being a
fake, continuum background). We hope therefore, that the other
backgrounds will be similar to one or other of these in their
behaviour. The recent CMS results suggest, moreover, that our two
backgrounds are the dominant ones in the
signal region in most of the $\tau \tau$ sub-channels (along with pure
jets, which we expect to be similar to $W+$jets).

Finally, we have only considered the most obvious systematic effect,
namely the uncertainty associated with the normalization of the signal
and backgrounds. Nevertheless, our qualtitative argument that a better
mass reconstruction gives a better separation between the signal and the
different backgrounds, means that many systematic uncertainties are expected
to be reduced.

\begin{figure}
\centering
\begin{minipage}[b]{0.45\linewidth}
\centering
\includegraphics[width=0.95\textwidth]{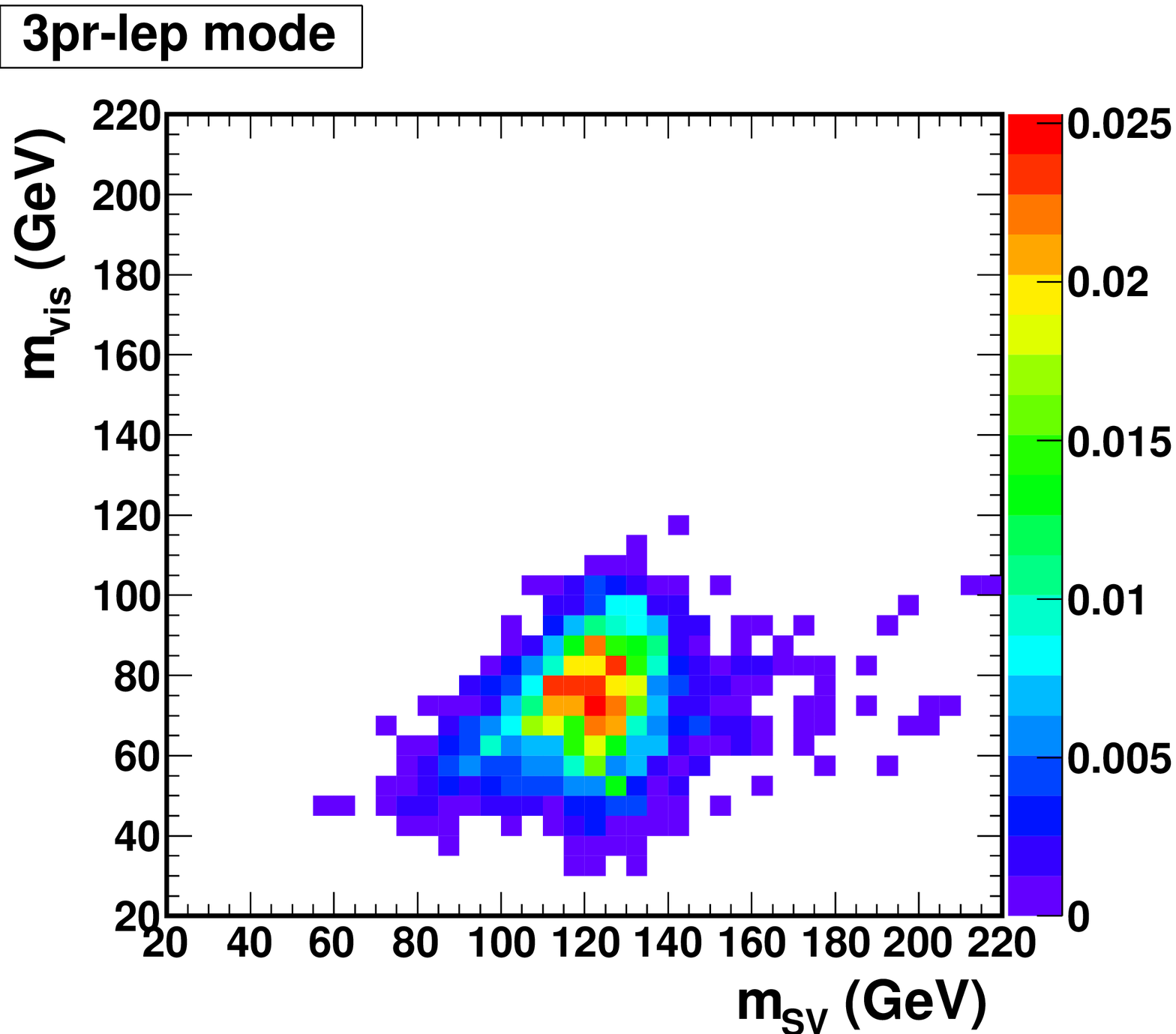}
\includegraphics[width=0.95\textwidth]{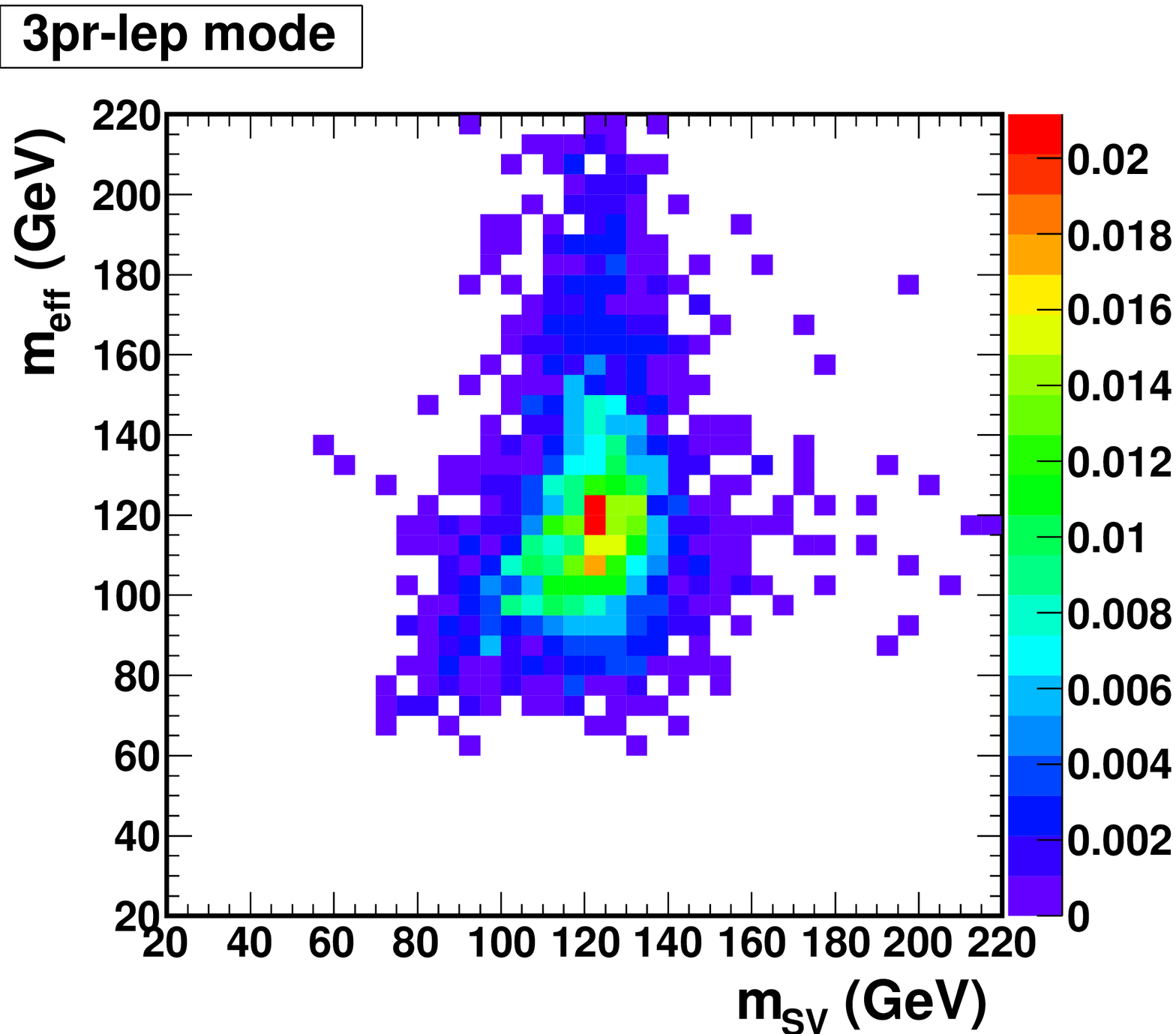}
\includegraphics[width=0.95\textwidth]{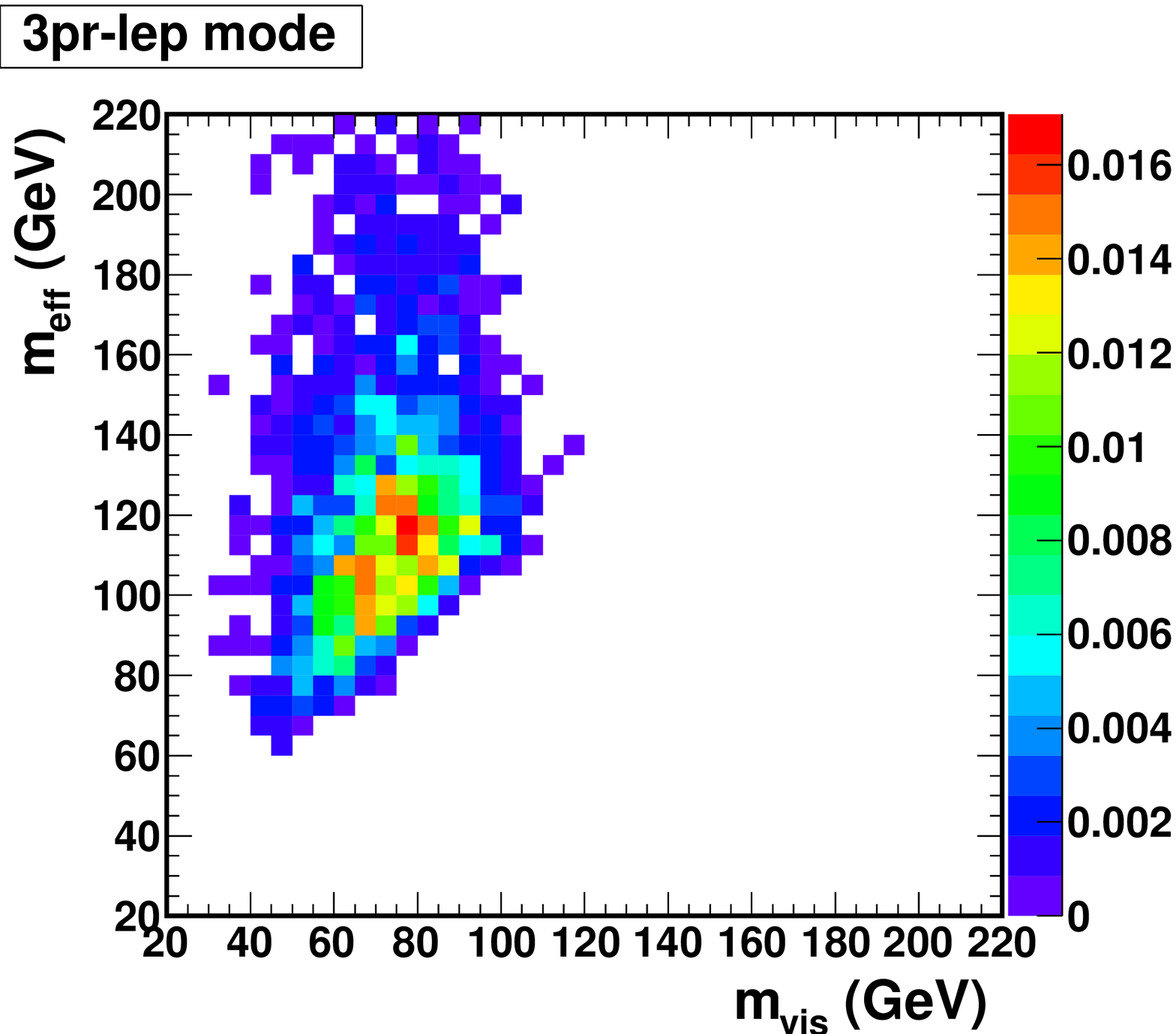}
\end{minipage}
\hspace{3mm}
\begin{minipage}[b]{0.45\linewidth}
\centering
\includegraphics[width=0.95\textwidth]{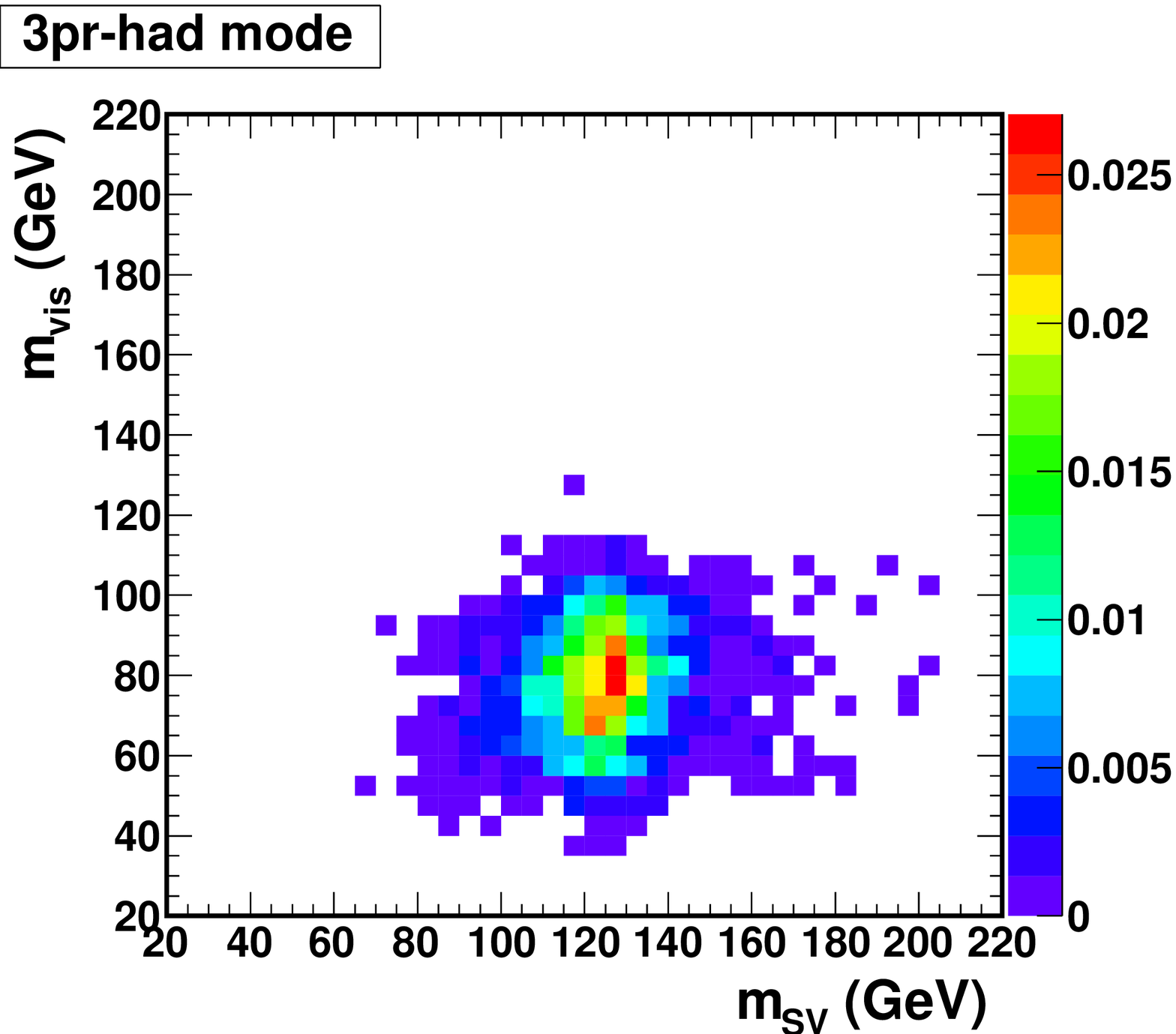}
\includegraphics[width=0.95\textwidth]{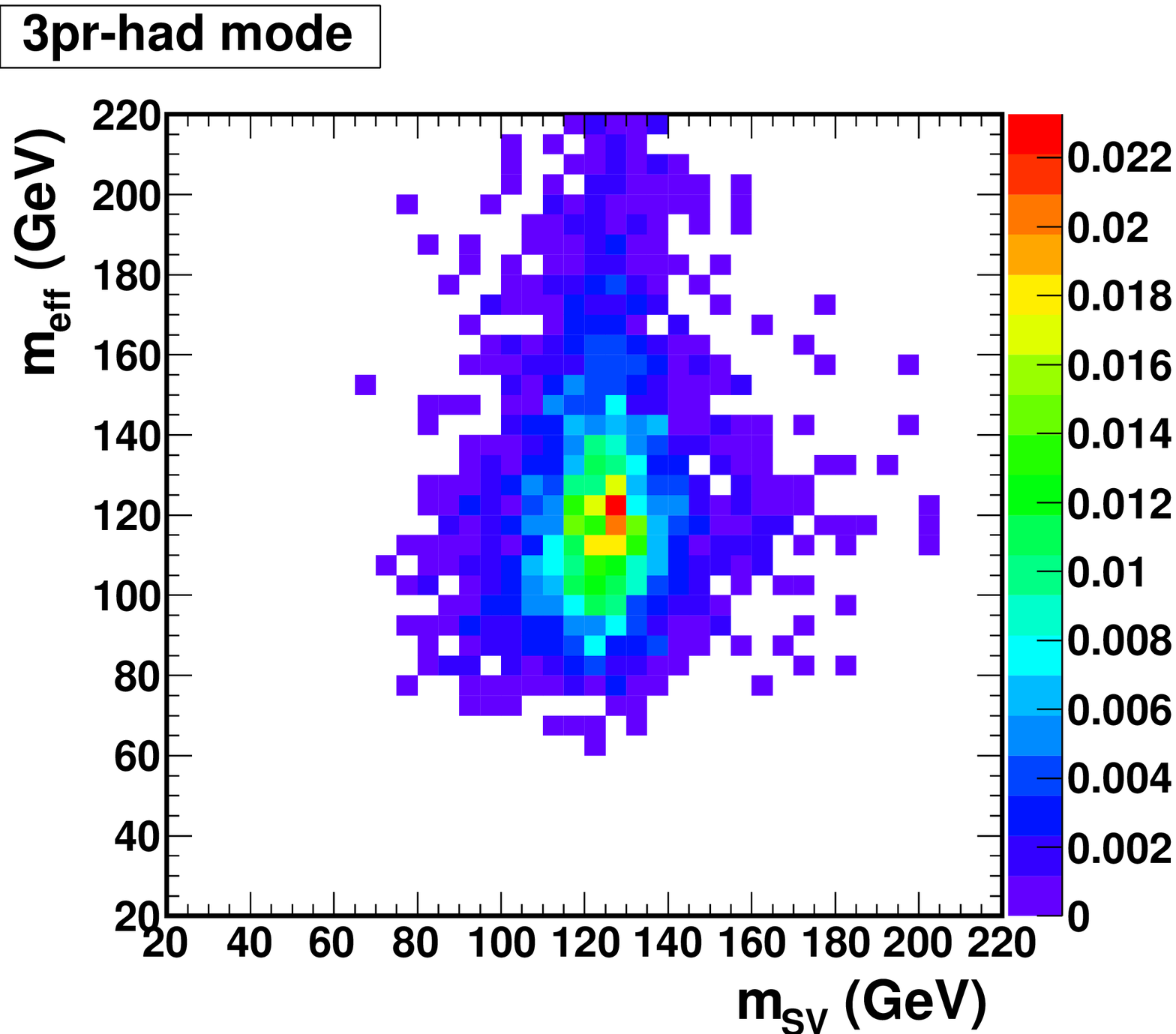}
\includegraphics[width=0.95\textwidth]{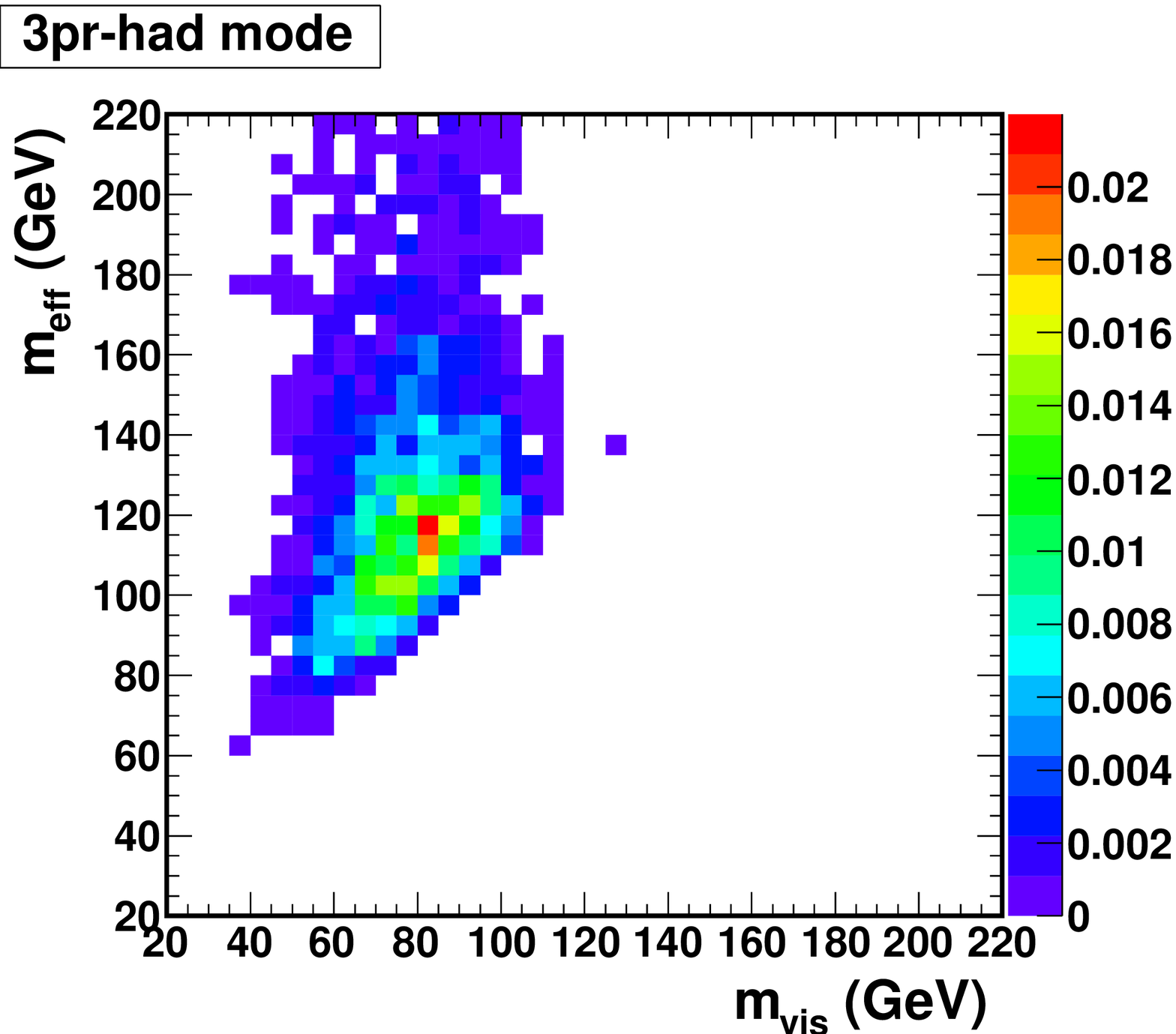}
\end{minipage}
\caption{Correlation between $m_{\mathrm{SV}}$,
  $m_{\mathrm{eff}}$ and  $m_{\mathrm{vis}}$, for the Higgs signal, for lepton-hadron (left)
  and hadron-hadron (right) modes.\label{fig:correlation}}
\end{figure}

We hope, at least, that our qualitative arguments and quantitative
simulations are enough to convince the collaborations to explore the
suitability of this method.
Even if it then turns out that a significant improvement is not obtained using our method
alone, we remark that an overall improvement can still be expected if it is combined
with existing approaches. Our method is complementary and, as is clear
from Fig~\ref{fig:correlation}, the observable we extract is not
strongly correlated with the existing variables  $m_{\mathrm{eff}}$
and  $m_{\mathrm{vis}}$.
It thus provides independent information and may be used to increase
the significance of searches in the $\tau \tau$ channel.
Our simulations apply to the current 8 TeV run and our hope is that
application of our method to data being produced now will allow us to
clear up the mystery of the observed deficit of $h \rightarrow \tau
\tau$ decays. Nevertheless, we expect our
method to become even more relevant in the subsequent stages of the
LHC programme. For one thing, the method is limited by statistics,
but gains in reducing systematic uncertainties. It is the
latter that will
limit our ultimate ability to make precision measurements of the Higgs
sector at the LHC. What is more, both
ATLAS and CMS are planning upgrades or replacements of their vertex detectors, with an
improvement of a factor of a few expected in the vertex
resolution. The associated improvement in the $\tau \tau$ mass
reconstruction using our method should reduce the uncertainties even further.

\section*{Acknowledgments}
BMG thanks A.~Barr, M.~Klute,  M.~Mulders, A.~de Roeck and the
Cambridge SUSY working group for discussions. He also acknowledges the
support of King's College, Cambridge and thanks Perimeter Institute
and the CERN Theory Group for hospitality.
This work is in part supported by Grant-in-Aid for Scientific research from the Ministry of Education, Science, Sports, and Culture (MEXT), Japan (Nos. 22540300, 23104005 for MN) and World Premier International Research Center Initiative (WPI Initiative), MEXT, Japan.
KS thanks K. Rolbiecki for helpful discussions.
BW acknowledges the support of a Leverhulme Trust Emeritus Fellowship,
and thanks IPMU, the CERN Theory Group, the CCPP at New York
University, the Galileo Galilei Institute and the Pauli Institute at ETH/University of Zurich
for hospitality.  He also thanks the INFN and the Pauli Institute for support during
parts of this work.

\appendix
\section{The treatment of jet masses}

\begin{figure}[t!]
\centering
\includegraphics[ width=0.6\textwidth]{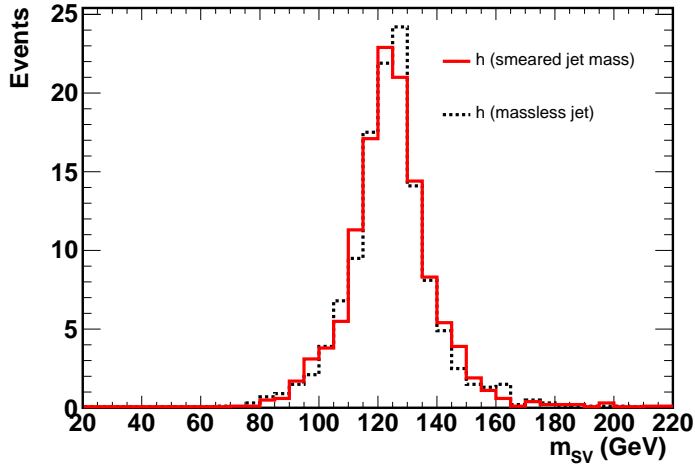}
\caption{
Comparison between two treatments of jet masses.
The red solid histogram uses jet masses smeared by the Gaussian distributions, whilst the black dashed histogram uses
massless jets.
The $h \to \tau\tau$ events in the hadron-hadron mode are used. 
\label{fig:jetmass}}
\end{figure}

As we discussed in section \ref{sec:method}, in evaluating the likelihood function (\ref{eq:like}),
we generate a large number of points ${\bf \tilde q}$ in which the jet energy is smeared, as well as the other observables, 
according to the detector resolution.
The magnitude of the jet momentum is then calculated as $p_j = \sqrt{E_j^2 - m_j^2}$, where $m_j$ is the jet mass.
We generate the jet mass according to the following probabilities:
\beq
P_{\rm 1 pr}(m_j) = G (m_j;m_\rho,\sigma_\rho) \cdot (1 - R_{\tau \to \pi \nu})   + \delta(m_j - m_\pi) \cdot R_{\tau \to \pi \nu},
\eeq
for the 1-prong tau-jet and
\beq
P_{\rm 3 pr}(m_j) = G (m_j;m_a,\sigma_a),
\eeq
for the 3-prong tau-jet, where $R_{\tau \to \pi \nu}$ is the $BR(\tau \to \pi \nu)$ divided by the branching ratio of inclusive 1-prong tau decays
and $G(x;\mu,\sigma)$ is Gaussian probability distribution with mean value $\mu$ and standard deviation $\sigma$.   
We took $m_{\rho} = 775$\,MeV, $\sigma_\rho = 90$\,MeV, $m_{a} = 1230$\,MeV and $\sigma_a = 160$\,MeV
in our analysis.

We have checked that 
the treatment of the jet mass does affect the frequency of finding real solutions in the likelihood evaluation, but does not
affect much the overall shape of the likelihood function.  
Thus, $m_{\rm SV}$ is rather robust against variation of the jet mass.
To demonstrate this, we show two $m_{\rm SV}$ distributions for the $h \to \tau \tau$ events in the hadron-hadron channel in 
Figure \ref{fig:jetmass},
where one uses the jet mass described above and the other uses massless jets, with $P_{\rm 1 pr}(m_j) = P_{\rm 3 pr}(m_j) \propto \delta(m_j)$. 
As can be seen, the two distributions are very similar.

\bibliography{htautau}
\bibliographystyle{utphys}
\end{thebibliography}

\end{document}